\documentclass[twocolumn,showpacs,preprintnumbers,amsmath,amssymb]{revtex4}

\usepackage{rotating}

\usepackage{graphicx,epsfig}% Include figure files
\usepackage{dcolumn}% Align table columns on decimal point
\usepackage{bm}% bold math
\usepackage{color}

\def\cH{{\cal H}}

\def\tE{{\tilde E}}

\newcommand{\req}[1]{Eq.~(\ref{#1})}
\newcommand{\avg}[1]{\langle #1\rangle}

\newcommand{\fig}[1]{Fig.~\ref{#1}}

\DeclareMathOperator*{\argmin}{{\rm argmin}}	

\newcommand{\cut}[1]{{}}
\newcommand{\etal}[1]{\emph{~et al.}}

\newcommand{\iit}{i_t}
\newcommand{\itp}{i_{t+1}}
\newcommand{\itm}{i_{t-1}}
\newcommand{\jt}{j_t}
\newcommand{\jtp}{j_{t+1}}
\newcommand{\jtm}{j_{t-1}}
\newcommand{\ktp}{k_{t+1}}
\newcommand{\ltp}{l_{t+1}}
\newcommand{\ktm}{k_{t-1}}
\newcommand{\ijt}{{ij, t}}
\newcommand{\jit}{{ji, t}}

\newcommand{\jitm}{{ji, t-1}}
\newcommand{\kit}{{ki, t}}
\newcommand{\kitm}{{ki, t-1}}

\newcommand{\esp}{E^{st}}
\newcommand{\etsp}{\tilde{E}^{st}}
\newcommand{\ijtf}{{\iit\to\jtp}}
\newcommand{\ijtb}{{\iit\to\jtm}}
\newcommand{\kitf}{{\ktm\to\iit}}
\newcommand{\kitbb}{{\ktp\to\iit}}
\newcommand{\jitf}{{\jtm\to\iit}}
\newcommand{\jitb}{{\jtp\to\iit}}
\newcommand{\litb}{{\ltp\to\iit}}
\newcommand{\sijtmu}{\sigma_\ijt^\nu}
\newcommand{\sjitmu}{\sigma_\jit^\nu}

\newcommand{\Ltmu}{\Lambda^\nu_{t}}

\newcommand{\red}{\textcolor{black}}

\begin{document}

\preprint{}

\title[Title]
{Coordinating Dynamical Routes with Statistical Physics on Space-time Networks}
\author{Chi Ho Yeung}
\affiliation{Department of Science and Environmental Studies, The Education University of Hong Kong, Tai Po, Hong Kong, China}

\date{\today}

\begin{abstract}
Coordination of dynamical routes can alleviate traffic congestion and is essential for the coming era of autonomous self-driving cars. However, dynamical route coordination is difficult and many existing routing protocols are either static or without inter-vehicle coordination. In this paper, we first apply the cavity approach in statistical physics to derive the theoretical behavior and an optimization algorithm for dynamical route coordination, but they become computational intractable as the number of time segments increases. We therefore map static spatial networks to space-time networks to derive a computational feasible message-passing algorithm compatible with arbitrary system parameters; it agrees well with the analytical and algorithmic results of conventional cavity approach and outperforms multi-start greedy search in saving total travel time by as much as $15\%$ in simulations. The study sheds light on the design of dynamical route coordination protocols, and the solution to other dynamical problems via static analytical approaches on space-time networks.
\end{abstract}

\pacs{89.75.Hc, 02.50.-r, 05.20.-y, 89.20.-a}
% 89.75.Hc: Networks and genealogical trees 
% 02.50.-r: Probability theory, stochastic processes, and statistics
% 05.20.-y: Classical statistical mechanics
% 89.20.-a: Interdisciplinary applications of physics

\maketitle

%%%%%%%%%%%%%%%%%%%%%%%%%%%%%%%%%%%%%%%%%%%%%%

\section{Introduction}

Traffic congestions occur everywhere in the world, especially in metropolitan areas where road expansion is not feasible~\cite{mogridge1997self}. To alleviate congestion, it is essential to maximize the traffic capacity of the fixed infrastructure by coordinating traffic flows. For instance, Beijing and Singapore implement driving restriction by license plate number~\cite{sun2014restricting} and electronic road pricing system~\cite{goh2002congestion} respectively, but these centralized regulations may not be applicable to every city. Dynamic traffic assignment (DTA)~\cite{merchant1978model} was proposed to coordinate dynamical traffic routes, with tools such as linear programming which are not favorably scalable and inadaptive to incremental changes~\cite{peeta2001foundations}. Nevertheless, the success of dynamical route coordination would not only alleviate congestion, but is also essential for coordinating routes in the coming era of autonomous self-driving cars~\cite{gerla2014internet}.

Physicists have devoted extensive efforts to understand and derive applications for transportation networks. For instance, models are built to understand the growth of road networks~\cite{lammer2006scaling, wu2006transport}; traffic dynamics are modeled by cellular automata~\cite{nagel1992cellular}, diffusion~\cite{gomez2013diffusion}, random walks~\cite{wang2006traffic} and user equilibrium~\cite{youn2008price}. Principled statistical physics tools are applied to optimize transportation networks~\cite{yeung12, yeung13, de2014shortest, altarelli2015edge}. These fundamental understandings always lead to applications to improve transportation networks. Nevertheless, the improvement strategies derived are either heuristic or applicable only on static path assignment. A way to coordinate dynamical traffic flows is still a great challenge to the physics community. 

In this paper, we apply the cavity approach~\cite{mezard87} to establish a theoretical framework to analyze dynamical route coordination. Theoretical behavior and an optimization algorithm are derived, but are limited by the computational complexity when the number of time segments increases. We therefore map the spatial networks to space-time (ST) networks~\cite{zawack1987dynamic}, and derive a message-passing algorithm capable to coordinate dynamical route of multiple vehicles, compatible with a large number of time segments. Despite the presence of structured short loops, the ST algorithm agrees well with the analytical and algorithmic results of cavity approach and outperforms multi-start greedy (MSG) search~\cite{chen1996new, srinivas2003minimum} in saving total traveling time by as much as $15\%$ in simulations. The study shed light on the design of dynamical route coordination protocols, and the solution to other dynamical problems via static analytical approaches on space-time networks.

\section{Problem formulation}
\label{sec_model}

We consider $M$ vehicles, denoted by $\nu=1,\dots,M$, traveling on a transportation network in a period from time $t\!=\!0$ to $T$. The network has $N$ nodes, each represents a site and is connected (e.g. by roads) to $K_i$ neighboring sites, with $a_{ij}\red{=a_{ji}}\!=\!1$ if node $i$ and $j$ are connected, and otherwise $a_{ij}\red{=a_{ji}}\!=\!0$. Each vehicle $\nu$ starts from an origin node $O_\nu$ at time $t_\nu$ and travels to a destination node $D_\nu$. We denote the variable $\sijtmu\!=\!1$ if vehicle $\nu$ passes the \red{directed link $ij$ from node $i$ to node $j$ between time $t$ and $t\!+\!1$}\red{, and otherwise $\sijtmu\!=\!0$. Here, we assume that each connection between neighboring nodes is bi-directional such that $\sijtmu$ and $\sjitmu$ are independent variables}. 
%Since time always goes forward, $\sijtmu\in \{0,1\}$ only. If we assume each link can be occupied by at most one vehicle, as restricted by Eq.~\req{eq_con4}, between $\sijtmu$ and $\sjitmu$, at most one of them is non-zero since they are the same link
Our goal is to identify the dynamical paths for all vehicles minimizing the total cost (or total travel time):
\begin{align}
\label{eq_H}
\cH = \sum_{i,j=1}^{N}\sum_{t=0}^{T-1}\sum_{\nu=1}^{M} a_{ij}\sigma_{\ijt}^\nu [1+\delta_{ij}(w_s-1)],
\end{align}
where Kronecker delta $\delta_{ij}=1$ when $i=j$ and otherwise $\delta_{ij}=0$, and we consider $a_{ii}=1$ for all $i$ such that the cost $\cH$ increases by 1 when a vehicle travels between two nodes, or by $w_s$ when it waits at a node for one step. 

\begin{figure}
%\centerline{\epsfig{figure=bidirectional.eps, width=0.7\linewidth}}
\centerline{\includegraphics[width=0.7\linewidth]{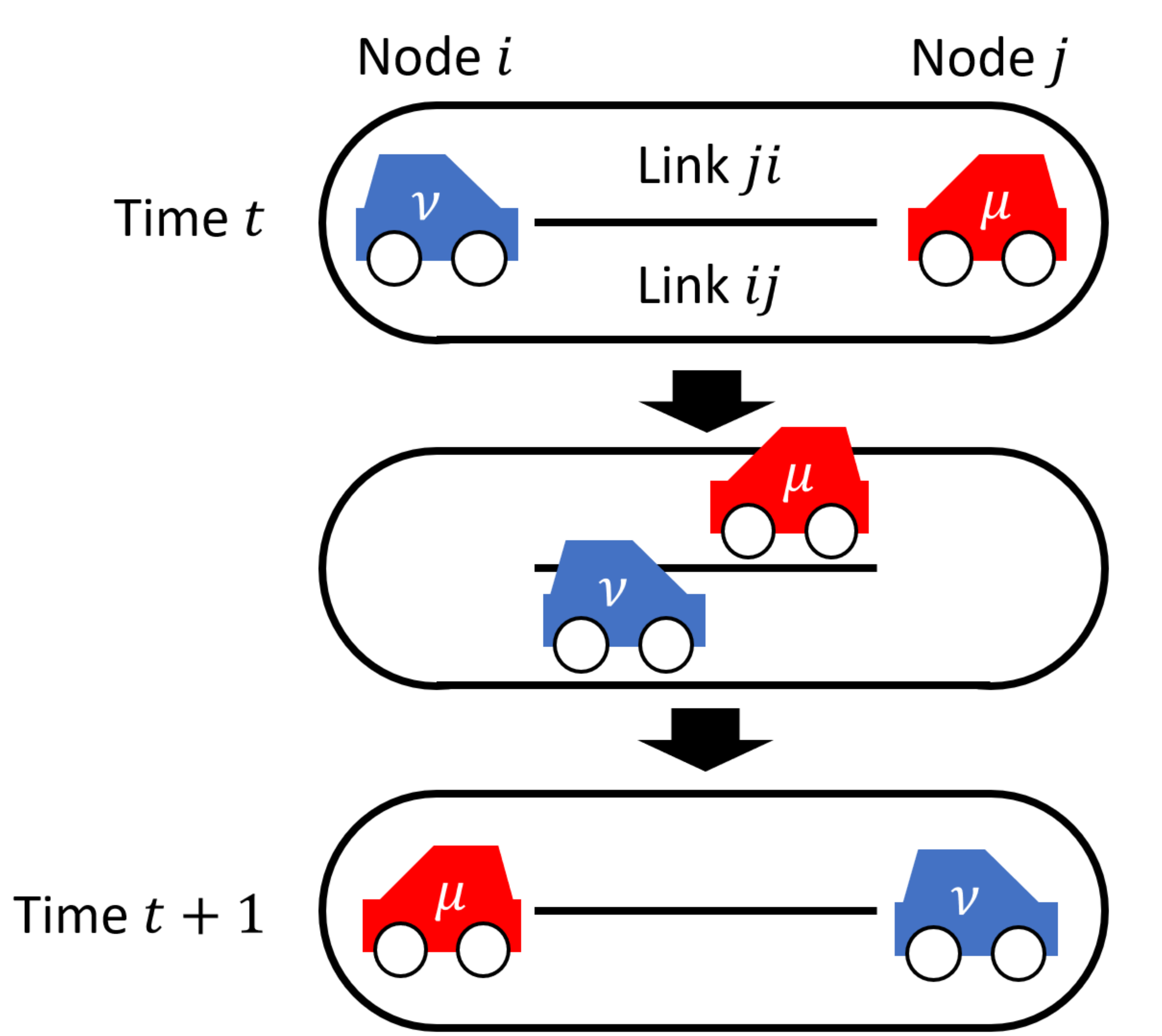}}
\caption{\red{An example to illustrate a bi-directional connection between node $i$ and $j$ in the network. At time $t$, vehicles $\nu$ and $\mu$ occupy node $i$ and node $j$ respectively. Vehicle $\nu$ then travels from node $i$ to $j$ via the directed link $ij$ from time $t$ to $t+1$, while vehicle $\mu$ travels in the opposite direction from node $j$ to $i$ via the directed link $ji$ at the same time.}}
\label{fig_bidirection}
\end{figure}

To reveal the typical behavior of the system, we map it to a problem of resource allocation and define the resource $\Lambda^\nu_{i,t}$ on node $i$ at time $t$ for vehicle $\nu$ to be
\begin{align}
\Lambda^\nu_{i,t}=
\begin{cases}
-1 & \mbox{if}\quad i=O^\nu, \quad t=t^\nu
\\
1 & \mbox{if}\quad i=D^\nu, \quad\forall t
\\
0 & \mbox{otherwise}
\end{cases}
\end{align}
Identifying paths for vehicle $\nu$ is equivalent to conserving the resources of $\nu$ on all nodes and time segments by
\begin{align}
\label{eq_con2}
\Lambda^\nu_{i,t} + \red{\sum_{j=1}^N a_{ij}\left[\sigma_{\ijt}^\nu - \sigma_{\jitm}^\nu\right]} = 0, \quad\forall i, \nu, t
\end{align}
We further assume each node and each \red{directed} link can be occupied by at most one vehicle at time $t$ \red{(for cases where nodes and links can be occupied by more than one vehicles, e.g. roads with multiple lanes, see Appendix~\ref{sec_lane})}, given by
\begin{align}
\label{eq_con3}
\sum_{\nu=1}^M\sum_{j=1}^N a_{ij}\left(\red{\sigma_\jitm^\nu+\sigma_\ijt^\nu}\right) \le 2, \quad\forall i, t
\\
\label{eq_con4}
\sum_{\nu=1}^M a_{ij}\red{\sigma_\ijt^\nu} \le 1, \quad\forall i,j,t.
\end{align}
which are constraints similar to those in the node-disjoint and edge-disjoint path problems~\cite{de2014shortest, altarelli2015edge}. \red{Since we consider each connection between neighboring nodes in the network is bi-directional, the constraint \req{eq_con4} applies independently for the two directed links $ij$ and $ji$, and it is possible for both $\sigma_\ijt^\nu=1$ and $\sigma_\jit^\mu=1$ for some $\nu\neq\mu$, as shown in the example in \fig{fig_bidirection}.}

%%%%%%%%%%%% fig 1 %%%%%%%%%%%%
%%%%%%%%%%%%%%%%%%%%%%%%%%%

\section{Analytical Solution}
\label{sec_analytical}

We assume only large loops exist in the network and employ the zero-temperature cavity approach~\cite{mezard87} developed in the studies of spin glass to analyze the problem. We first define the energy of a tree terminated at node $i$ in the absence of neighbor $j$ to be $E_{i\to j}(\vec{\sigma}_{ij}, \vec{\sigma}_{ji})$, a function of flow variables $\vec{\sigma}_{ij}$ and  $\vec{\sigma}_{ji}$ on directed link $ij$ \red{and directed link $ji$ respectively}, where $\vec{\sigma}_{ij}\!=\!(\sigma^{\nu=1}_{ij, t=0},\dots,\sigma^{M}_{ij, 0}, \dots, \sigma^{1}_{ij, T},\dots,\sigma^{M}_{ij, T})$ and similarly for $\vec{\sigma}_{ji}$. We then express $E_{ij}(\vec{\sigma}_{ij}, \vec{\sigma}_{ji})$ in terms of $E_{ki}(\vec{\sigma}_{ki}, \vec{\sigma}_{ik})$ as
\begin{align}
\label{eq_iteration}
&E_{i\to j}(\vec{\sigma}_{ij}, \vec{\sigma}_{ji}) 
= \min_{\{\vec{\sigma}_{ii}, \{\vec{\sigma}_{ki}, \vec{\sigma}_{ik}\}|\{C_3^{i, \nu, t}, C_4^{i, t}, C_5^{ik, t}\}_{\forall k, \nu, t}\}}\Big[|\vec{\sigma}_{ij}|
\nonumber\\
&\quad\left.+|\vec{\sigma}_{ji}|+w_s|\vec{\sigma}_{ii}|+\sum_{k=1}^{K_i-1}E_{k\to i}(\vec{\sigma}_{ki}, \vec{\sigma}_{ik})\right]
\end{align}
where $C_3^{i, \nu, t}, C_4^{i, t}$ and $C_5^{ij, t}$ correspond to constraints in the form of \req{eq_con2} for node $i$ and vehicle $\nu$; \req{eq_con3} for node $i$; and \req{eq_con4} for directed link $ij$, all at time $t$. \red{We note that the energy function $E_{i\to j}(\vec{\sigma}_{ij}, \vec{\sigma}_{ji})$ plays the role of cavity fields in the conventional cavity approach. However, to facilitate the derivation of the recursion relation \req{eq_iteration}, we express the cavity energy function in terms of the flow variables $\vec{\sigma}_{ij}$ and  $\vec{\sigma}_{ji}$ despite the absence of neighbor $j$, as in other studies of flow optimization using the cavity approach~\cite{yeung12, yeung13, de2014shortest, yeung14, wong06, yeung09}. One can set  $\vec{\sigma}_{ij} = \vec{\sigma}_{ji} =\vec{0}$ in cases if the absence of neighbor $j$ has to be considered.}

%%%%%%%%%%%% fig 2 %%%%%%%%%%%%
\begin{figure}
\centerline{\includegraphics[width=0.9\linewidth]{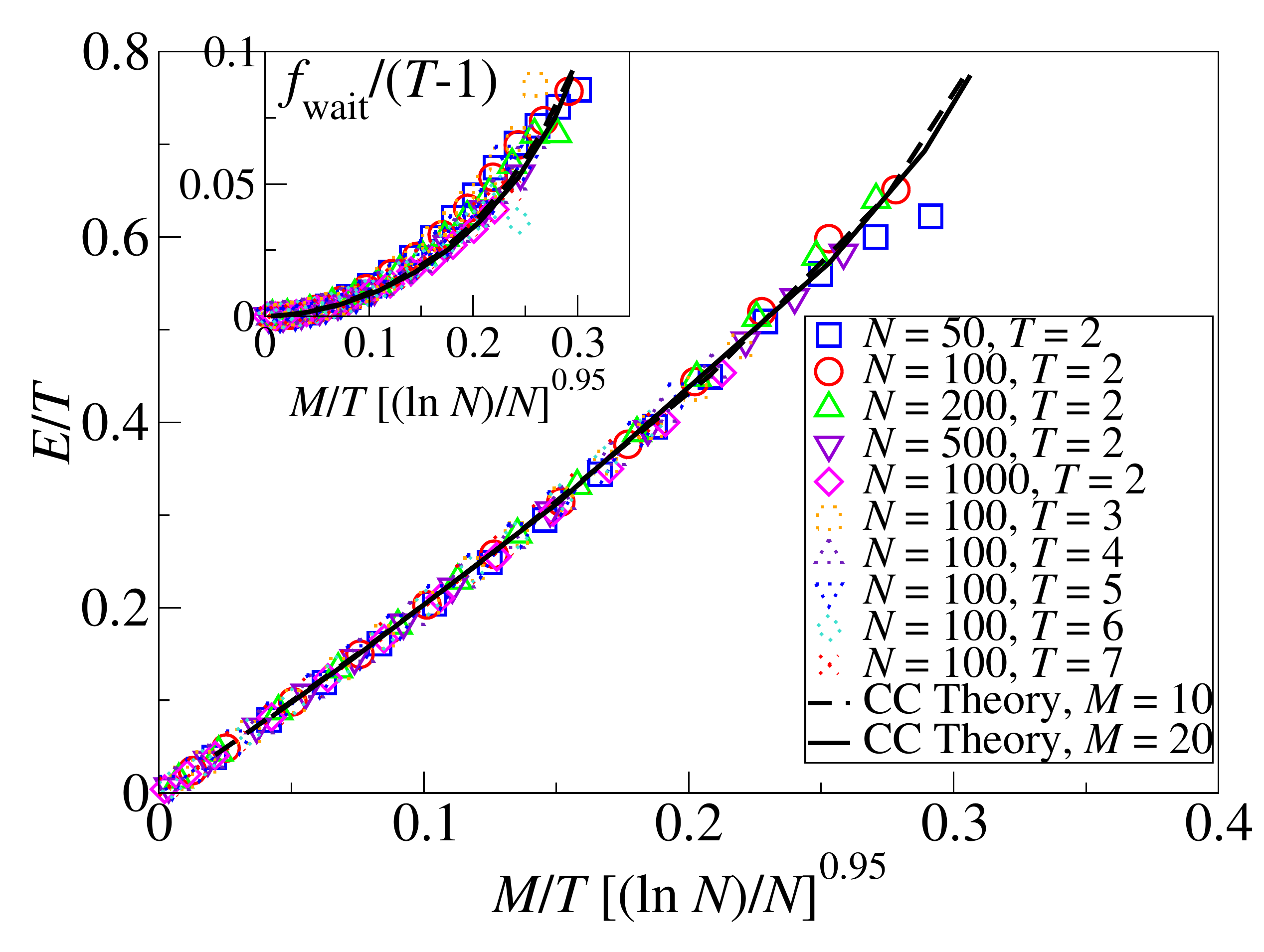}}
%\centerline{\includegraphics[width=0.9\linewidth]{energyf3.pdf}}
\caption{The energy $\avg{E}/T$ optimized by the ST algorithm as a function of $M/T (\ln N/N)^{0.95}$, for random regular graphs with $K=3$ and various $N$, $M$ and $T$. \red{The results are averaged over the converged realizations in at least 1000 instances, and the ST algorithm is terminated at a maximum of $5\times 10^4 N$ updates if the messages do not converge. The energy obtained by the ST algorithm on cases with various values of $K$ can be found in Appendix~\ref{sec_degree}.} Inset: The results of $\avg{f_{\rm wait}}$, i.e. the fraction of nodes with a waiting vehicle, as a function of $M/T (\ln N/N)^{0.95}$; $\avg{f_{\rm wait}}$ is scaled with $T-1$ since a vehicle can wait for at most $T-1$ time steps on a node. The theoretical results are obtained by solving \req{eq_iteration} with $T=2$, $M=10, 20$ and various $1/N$.}
\label{fig_energy}
\end{figure}
%%%%%%%%%%%%%%%%%%%%%%%%%%

Since $E(\vec{\sigma}_{ij}, \vec{\sigma}_{ji})$ is an extensive quantity, we define an intensive quantity
\begin{align}
\label{eq_tildeE}
\tE_{i\to j}(\vec{\sigma}_{ij}, \vec{\sigma}_{ji}) = E_{i\to j}(\vec{\sigma}_{ij}, \vec{\sigma}_{ji}) - E_{i\to j}(\vec{0}, \vec{0}).
\end{align}
By substituting \req{eq_iteration} into \req{eq_tildeE}, one can write down a recursion relation in terms of $\tE_{ij}$ instead of $E_{ij}$. To compute the quantites of interests, we employ population dynamics to iterate \req{eq_iteration} and obtain a steady distribution $P[\tE(\!\vec{\sigma}_{ij},\!\vec{\sigma}_{ji}\!)]$, given a node degree distribution $P(K)$ and the distribution of vehicle origin and destination
\begin{align}
\label{eq_lambda}
P(\vec{\Lambda})\!=\!\prod_{\nu=1}^{M}\prod_{t=0}^{T}\left[\frac{1}{N}\delta_{\Ltmu\!,1} \!+\!\frac{1}{N}\delta_{\Ltmu\!,-1}\!+\!\left(\!1\!-\!\frac{2}{N}\!\right)\delta_{\Ltmu,0}\right]
\end{align}
Since the cavity approach assumes an infinite system size, the density of vehicle would tend to zero if $M$ remains finite. Instead of using an infinite $M$, we assume multiple origins and destinations for $\nu$ such that $1/N$ in \req{eq_lambda} becomes a parameter to characterize the origin and destination density, instead of system size. After convergence of $P[\tE(\vec{\sigma}_{ij}, \vec{\sigma}_{ji})]$, we compute the average cost of adding a node and a link by 
\begin{align}
\label{eq_enode}
&E_{\rm node}=
\\
&\left\langle\!\min_{\{\vec{\sigma}_{ii}, \{\vec{\sigma}_{ki}, \vec{\sigma}_{ik}\}|\{C_3^{i,\nu,t},C_4^{i,t}\}_{\forall \nu, t}\}}\!\left[
w_s|\vec{\sigma}_{ii}|\!+\!\sum_{k=1}^{K_i}\tE_{k}(\vec{\sigma}_{ki}, \vec{\sigma}_{ik})\right]\!\right\rangle
%_{\!\{\!E_{k}\!\},\Lambda_i,K_i}
\nonumber\\
\label{eq_elink}
&E_{\rm link}= 
\\
&\left\langle\!\min_{\{\vec{\sigma}_{ki}, \vec{\sigma}_{ik}|\{C_5^{ik, t}\}_{\forall t}\}}\!\Big[\tE_1(\vec{\sigma}_{ki}, \vec{\sigma}_{ik})\!+\!\tE_2(\vec{\sigma}_{ik}, \vec{\sigma}_{ki})\!-\!|\vec{\sigma}_{ki}|\!-\!|\vec{\sigma}_{ik}|\Big]\!\right\rangle
%_{\!E_1,E_2}
\nonumber
\end{align}
averaged over $P[\tE(\vec{\sigma}_{ik}, \vec{\sigma}_{ki})]$, $P(\vec{\Lambda}_i)$ and $P(K_i)$. The average cost per node is given by $\avg{E}=E_{\rm node} - \avg{K}E_{\rm link}/2$.

As constrained by \red{\req{eq_con3}, a node can be occupied by at most one vehicle at a time. We assume that a vehicle does not go back and forth on the same connection at consecutive time steps, e.g. travels from node $i$ to $j$ via the directed link $ij$ and then immediately travels back from node $j$ to $i$ via the directed link $ji$ at the next step. In this case,} there is at most one non-zero entry in 
%$(\sigma^{1}_{ij, t},\dots,\sigma^{M}_{ij, t})$ and $(\sigma^{1}_{ji, t},\dots,\sigma^{M}_{ji, t})$ 
either $\vec{\sigma}_{ij, t}$ or $\vec{\sigma}_{ji, \red{t-1}}$ at time $t$. Hence, the dimension of the domain of $E(\vec{\sigma}_{ij}, \vec{\sigma}_{ji})$ is $(2M+1)^T$ and the computational complexity of \req{eq_iteration} is $(2M+1)^{T(K_i-1)}$, which becomes intractable even for intermediate $M$, $T$ and $K_i$. To simplify the calculation, we assume a periodic time frame such that the time segments at $t\!=\!T$ and $t\!=\!0$ are the same, and paths at $t\!=\!T\!-\!1$ proceed to $t\!=\!0$, suitable for periodic routing and scheduling problems. We solved \req{eq_iteration} by population dynamics for $T=2$, with varying $1/N$ at each specific value of $M$; the results of $\avg{E}$ and $\avg{f_{\rm wait}}$, i.e. the fraction of nodes with a waiting vehicle (see Appendix~\ref{sec_fwait}), are shown in \fig{fig_energy}.

%To further reduce the complexity while keeping the dynamical nature of the problem, we only solve \req{eq_iteration} for $T=2$, and show by algorithmic results that systems with $T>2$ are dependent on $T$ only via the variable $M/T$.
%but will derive an algorithm for any $T>2$ in the next section and show that the physical quantities of the system depend on the rescaling parameter $M \log(N)/NT$ such that a system with large $T>2$ is equivalent to a system with $T=2$ but the same $M/T$.

%%%%%%%%%%%% fig 1 %%%%%%%%%%%%
\begin{figure}[t]
%\leftline{\mbox{(a)\hspace{3.8cm}(b)}}
%\vspace{-0.5cm}
%\centerline{\epsfig{figure=network2.eps, width=0.4\linewidth}
%\hspace{0.7cm}
%\epsfig{figure=spNet2.eps, width=0.4\linewidth}}
%\vspace{0.3cm}
%\leftline{\mbox{(c)\hspace{3.8cm}(d)}}
%\vspace{-0.5cm}
%\centerline{\epsfig{figure=spNetPeriodic2.eps, width=0.4\linewidth}
%\hspace{0.7cm}
%\epsfig{figure=spNetRoute2b.eps, width=0.4\linewidth}}
\leftline{\mbox{(a)\hspace{1.7cm}(b)\hspace{2.9cm}(c)}}
\vspace{-0.5cm}
\centerline{
\includegraphics[width=0.25\linewidth]{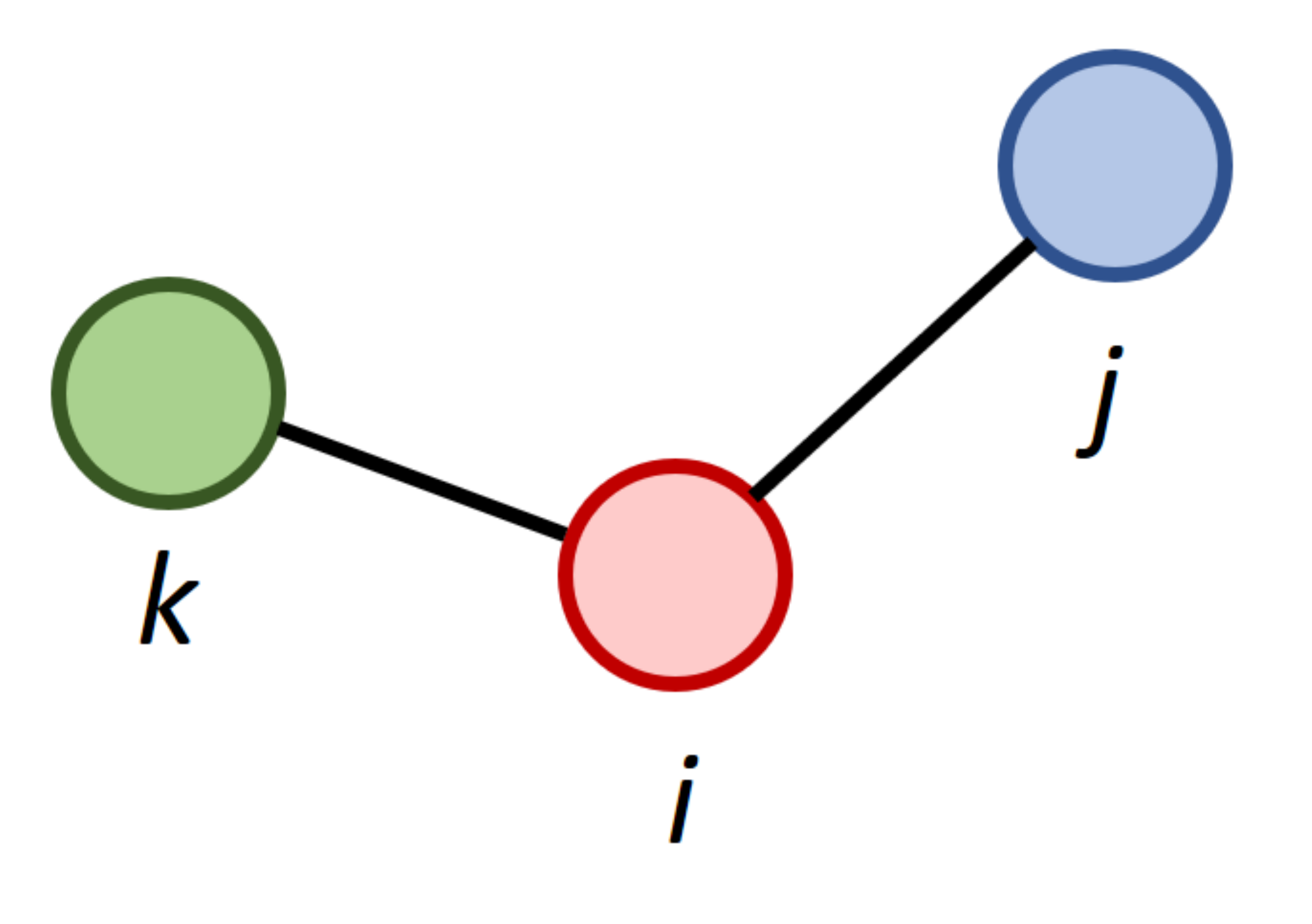}
\hspace{0.3cm}
\includegraphics[width=0.32\linewidth]{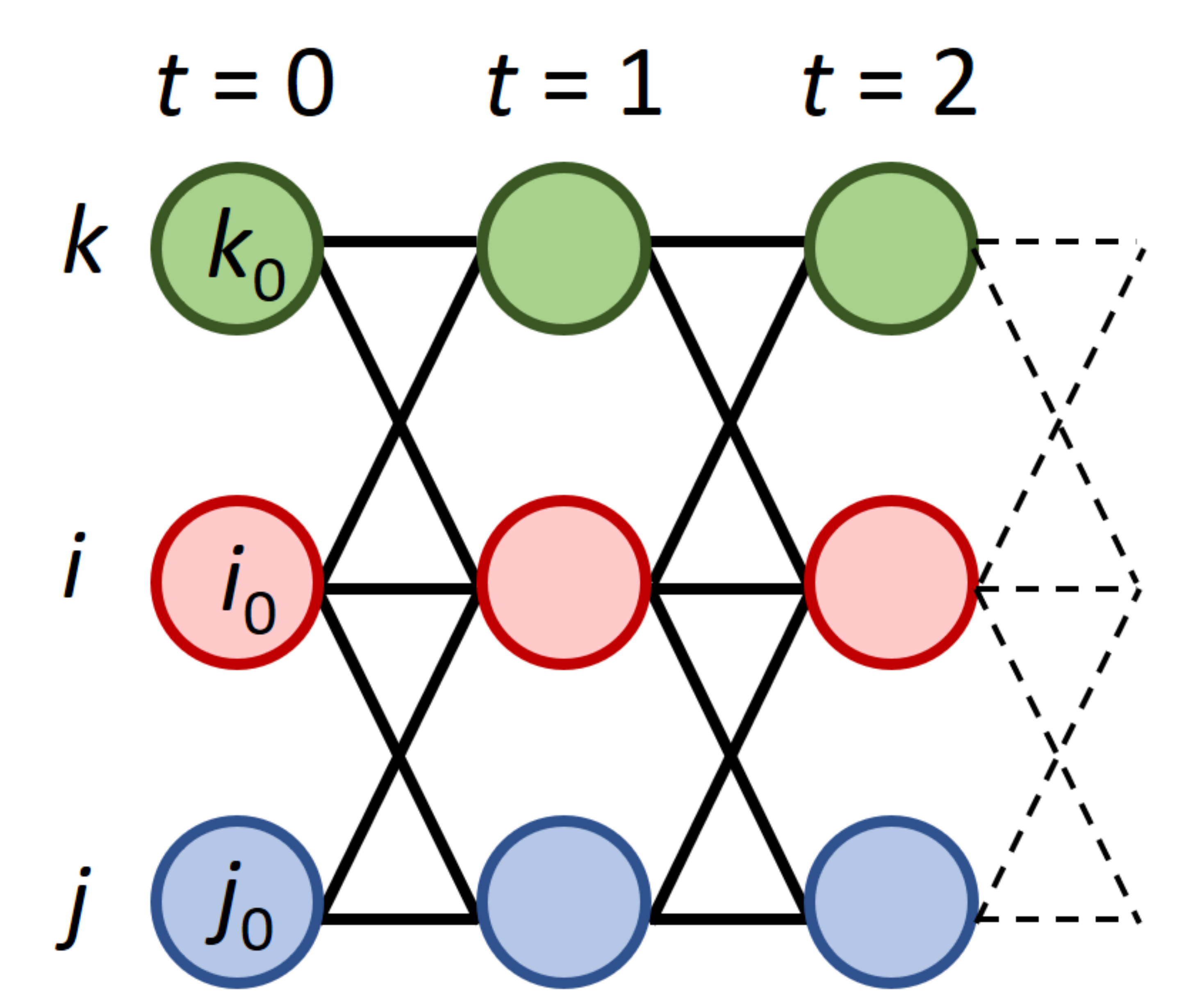}
\hspace{0.3cm}
\includegraphics[width=0.32\linewidth]{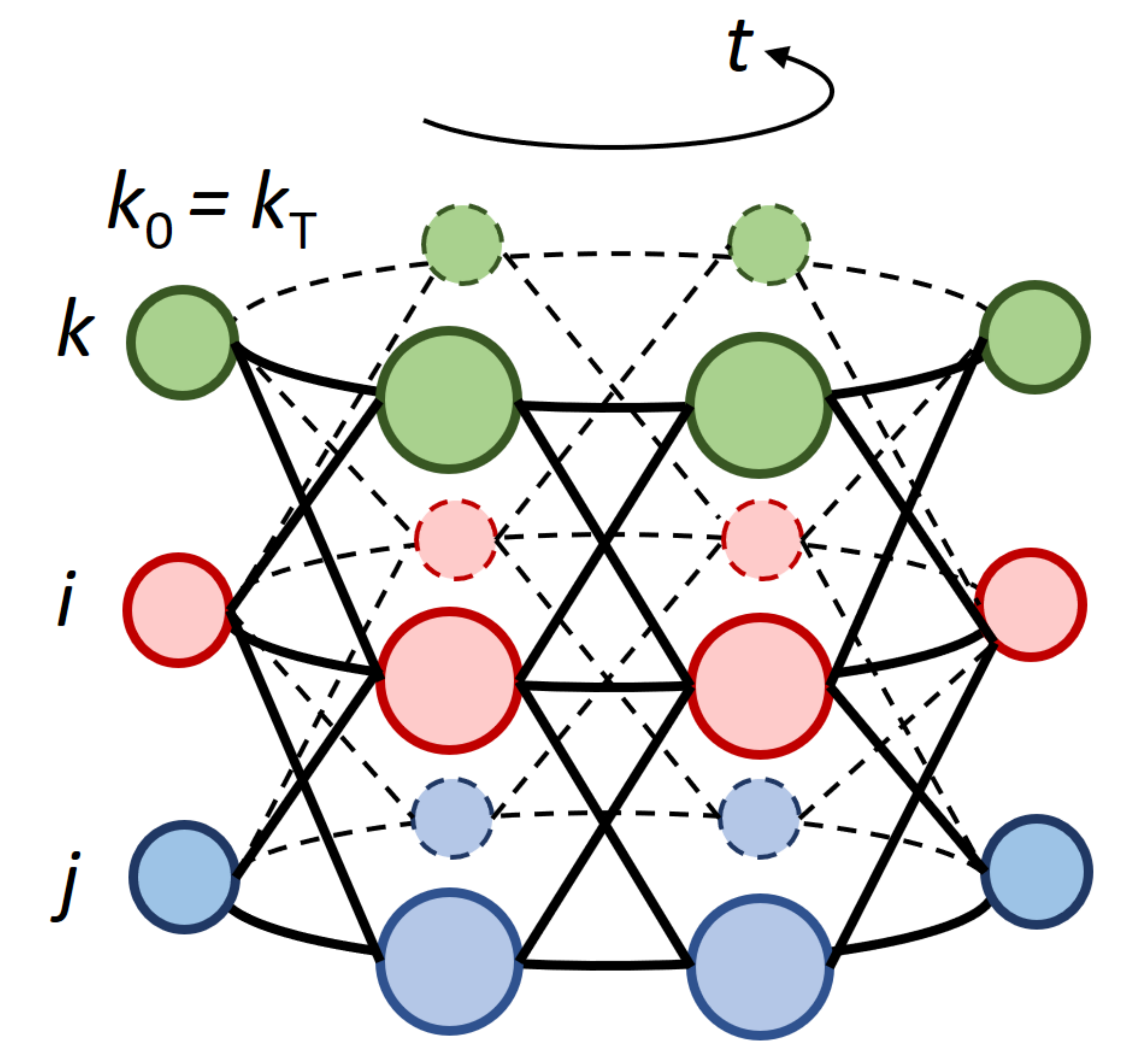}
}
\vspace{-0.2cm}
\caption{The example of spatial network shown in (a) is converted to a space-time network in (b) and a periodic space-time network in (c).}
\label{fig_spNet}
\end{figure}
%%%%%%%%%%%%%%%%%%%%%%%%%%

%%%%%%%%%%%% fig 3 %%%%%%%%%%%%
\begin{figure}
\centerline{\includegraphics[width=0.5\linewidth]{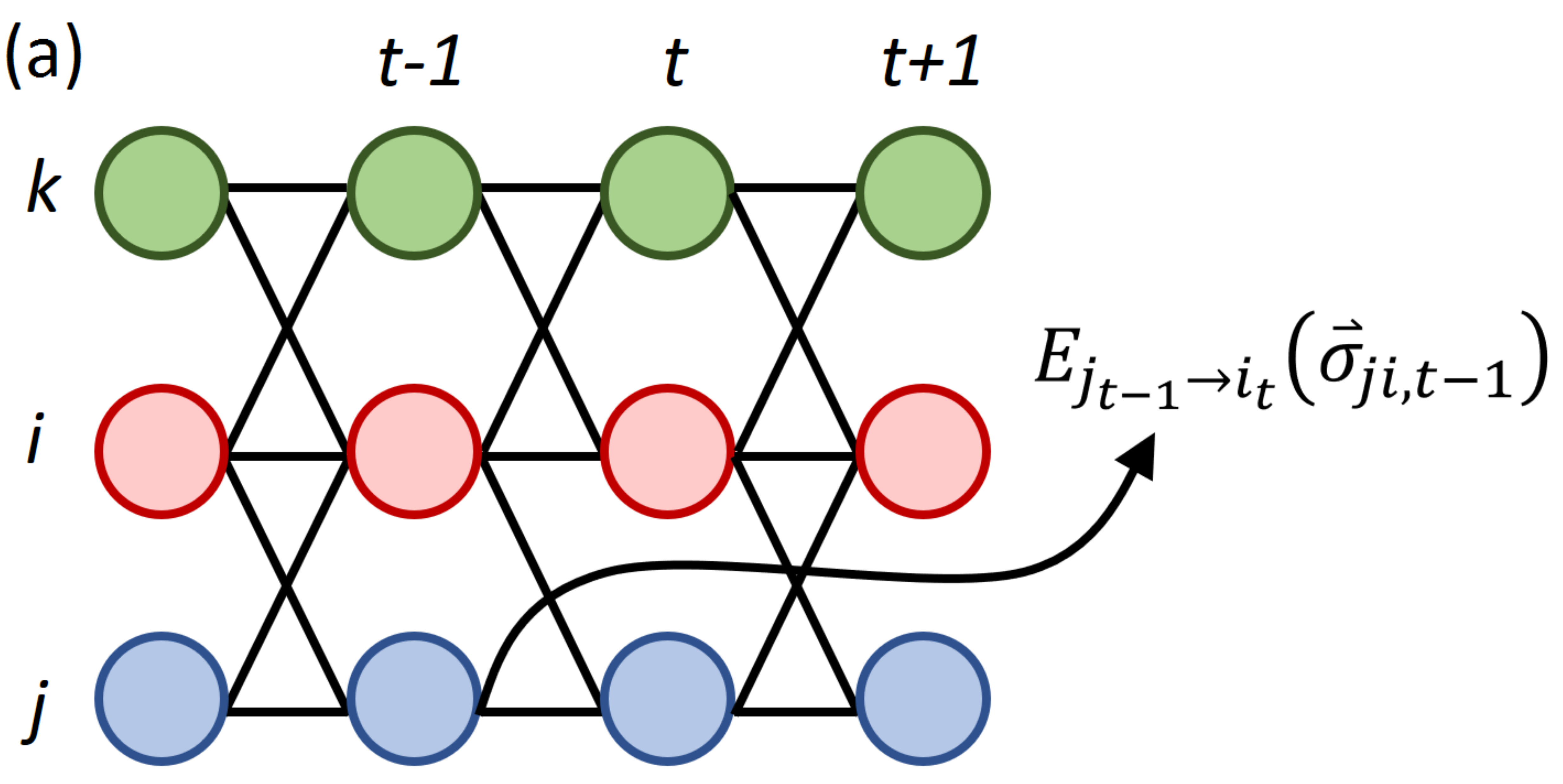}
\includegraphics[width=0.5\linewidth]{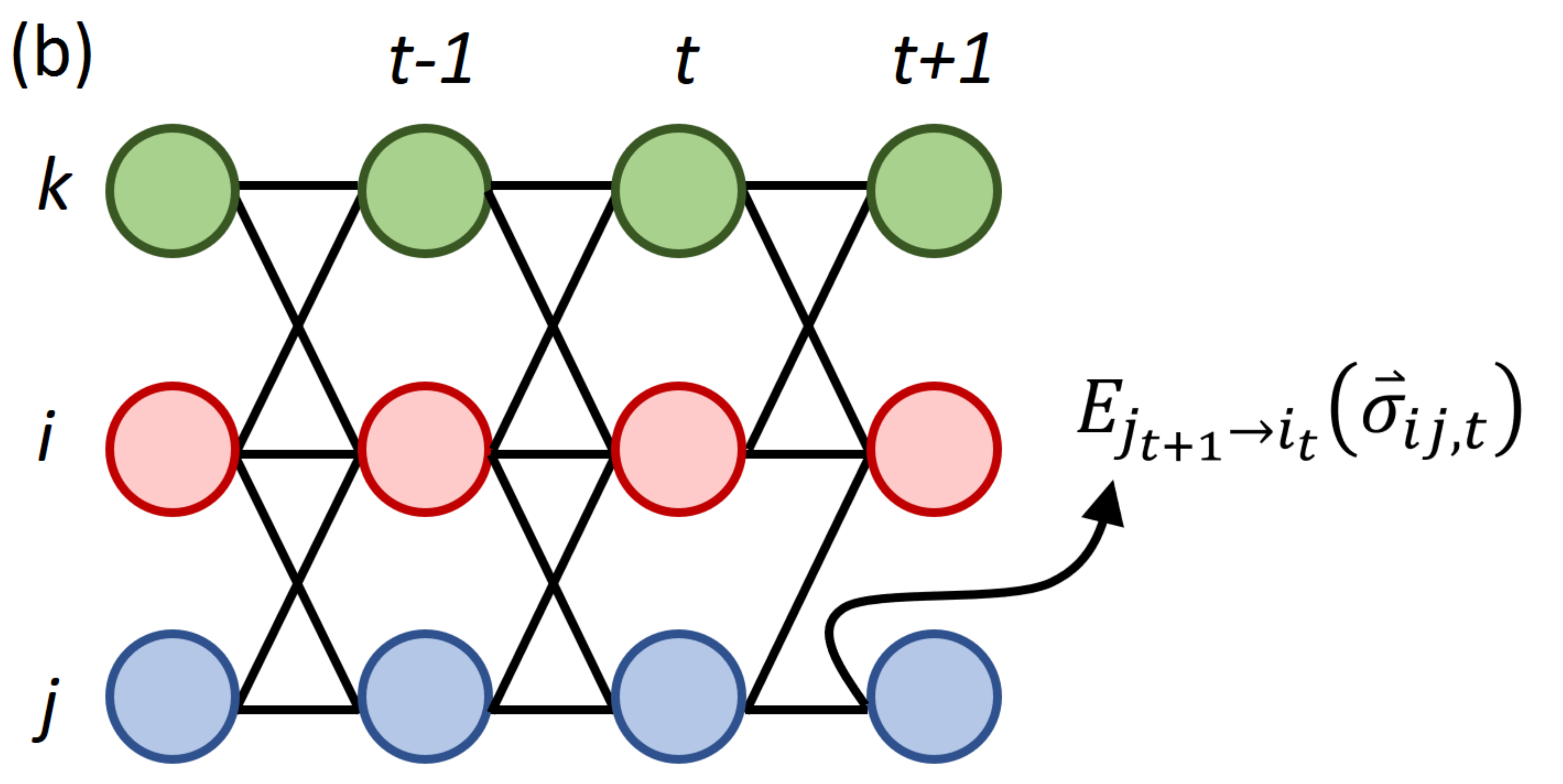}}
\caption{An example of (a) a forward energy function from $\jtm$ to $\iit$ and (b) a backward energy function from $\jtp$ to $\iit$.}
\label{fig_cavity}
\end{figure}
%%%%%%%%%%%%%%%%%%%%%%%%%%

\section{Optimization algorithm}
\label{sec_algorithm}

Equation (\ref{eq_iteration}) can also be used as an algorithm to optimize real instances, by passing $E_{ij}(\vec{\sigma}_{ij}, \vec{\sigma}_{ji})$ on all links of the network until convergence. We call this the \emph{conventional cavity} (CC) algorithm. Nevertheless, many real problems are characterized by $T\gg 2$, and the computational complexity of $O[(2M+1)^{T(K-1)}]$ makes \req{eq_iteration} intractable for many typical problems. To devise an algorithm with a feasible complexity, we convert the original network to a space-time (ST) network~\cite{zawack1987dynamic} as from \fig{fig_spNet} (a) to (b), where each node $\iit$ is characterized by one spatial coordinate and one time label. In the ST network, if node $i$ and $j$ are connected in the spatial network, node $\iit$ is connected to $\jtp$ and $\jtm$ instead of $\jt$. In addition, node $\iit$ is also connected to node $\itp$ and $\itm$. A path which passes between $\iit$ and $\itp$ corresponds to waiting at node $i$ between time $t$ and $t+1$.

When \req{eq_iteration} is used as an algorithm on a network, we assume only large loops exist such that the neighbors of a node are effectively independent when the node is removed. Since ST networks have many structured short loops, such assumption is greatly weakened. Nevertheless, the cavity approach has been successfully applied to lattice networks with numerous loops, especially on routing problems \cite{yeung14, altarelli2015edge}. We thus exploit this approximation and write down a cavity recursion relation in the ST network; we show later that the results of this ST algorithm agree well with the theory and algorithm of the CC approach derived in the absence of short loops.

As shown in \fig{fig_cavity}, we define an energy function $\esp_{\jitf}(\vec{\sigma}_{\jitm})$ on the \emph{forward} link from node $\jtm$ to node $\iit$, and similarly an energy function $\esp_{\jitb}(\vec{\sigma}_{ji, t-1})$ on the \emph{backward} link from $\jtp$ to $\iit$. We then write down the recursion relations for both forward and backward energy functions from node $\iit$
\begin{widetext}
\vspace{-0.4cm}
\begin{align}
\label{eq_forward}
\esp_{\ijtf}(\vec{\sigma}_{\ijt})
&\!=\hspace{-0.3cm}\min_{\substack{ \{\{\vec{\sigma}_{ki, t\!-\!1}\},\{\vec{\sigma}_{ik, t}\}_{k\neq j}|\vec{C}^t \}}}
\!\left[|\vec{\sigma}_{\ijt}|[1\!+\!\delta_{ij}(w_s\!-\!1)]
\!+\!\sum_k\!a_{ik}\esp_{k_{t\!-\!1}\!\to\iit}(\vec{\sigma}_{\kitm})\!+\!\sum_{k\neq j} a_{ik}\esp_{\kitbb}(\vec{\sigma}_{\kit})\!\right],
\\
\label{eq_backward}
\esp_{i_t\!\to\!j_{t\!-\!1}}(\vec{\sigma}_{ji, t\!-\!1})
&\!=\hspace{-0.3cm}\min_{\substack{ \{\{\vec{\sigma}_{ki,\!t-1}\}_{k\!\neq\!j},\{\vec{\sigma}_{ik, t}\}|\vec{C}^t  \}}}
\left[|\vec{\sigma}_{ji, t\!-\!1}|[1\!+\!\delta_{ij}(w_s\!-\!1)]
\!+\!\sum_{k\neq j} a_{ik}\esp_{k_{t\!-\!1}\!\to\iit}(\vec{\sigma}_{\kitm})\!+\!\sum_{k} a_{ik}\esp_{\kitbb}(\vec{\sigma}_{ik, t})\!\right]
\end{align}
\end{widetext}
where $\vec{C}^t$ represents constraints $\{C_3^{i, \nu, t}, C_4^{i, t}, C_5^{ik, t}\}_{\forall k, \nu}$. Since $\esp$ is an extensive quantity, one can again define an intensive quantity $\etsp$ as
\begin{align}
\label{eq_etspf}
\etsp_{\ijtf}(\vec{\sigma}_{\ijt}) = \esp_{\ijtf}(\vec{\sigma}_{\ijt}) - \etsp_{\ijtf}(\vec{0}),
\\
\label{eq_etspb}
\etsp_{\ijtb}(\vec{\sigma}_{\jitm})\!=\!\esp_{\ijtb}(\vec{\sigma}_{\jitm})\!-\!\etsp_{\ijtb}(\vec{0})
\end{align}
By substiting Eqs. (\ref{eq_forward}) and (\ref{eq_backward}) into Eqs. (\ref{eq_etspf}) and (\ref{eq_etspb}), we can write down the recursion relations in terms of $\etsp$ only. After convergence of messages $\etsp$ on all links, one can find the optimal flow on each directed link by
\begin{align}
\label{eq_config}
\vec{\sigma}_{\ijt}^*=\argmin_{\vec{\sigma}_{\ijt}}\left[\etsp_{\ijtf}(\vec{\sigma}_{\ijt})+\etsp_{\jtp\to\iit}(\vec{\sigma}_{\ijt})\!-\!|\vec{\sigma}_{\ijt}| \right],
\end{align}
which leads to a set of optimally-coordinated dynamical routes on the network. Due to constraints $C_3, C_4$ and $C_5$, the expression Eqs. (\ref{eq_forward}), (\ref{eq_backward}) and (\ref{eq_config}) can be greatly simplified, leading to a message passing algorithm derived in Appendix~\ref{sec_simSP}, which we call the \emph{space-time} (ST) algorithm. Hence, the dynamical route coordination problem is translated into a static ST network where static approaches can be applied.

We also remark that the dimension of the messages is greatly reduced from $(2M+1)^T$ in \req{eq_iteration} to $M$ in Eqs. (\ref{eq_forward}) and (\ref{eq_backward}), although the number of links in the ST network is $2T$ times more. The computation complexity of each update is also greatly reduced to $O(MKT)$. 

\red{The message-passing algorithm derived in Appendix~\ref{sec_simSP} is indeed equivalent to applying the node-disjoint path algorithm in \cite{de2014shortest} on space-time networks. Nevertheless, our goal to identify spatio-temporal path configurations is fundamentally different from that in \cite{de2014shortest}, as the later only identifies static path configurations without the temporal dimension. The presence of the time dimension leads to importance differences, e.g. simultaneous bi-directional traffic on a link is allowed in the present case as in \fig{fig_bidirection}, which has no equivalence in \cite{de2014shortest}. In addition, the ST networks studied here consist of numerous highly structured loops, and are different from the random regular graphs studied in \cite{de2014shortest} with only large loops. Investigating the validity of message-passing algorithms on ST graphs is one major contribution of the present work.}

\section{Results}

We show in \fig{fig_energy} and its inset the energy $\avg{E}/T$ and the fraction $\avg{f_{\rm wait}}/(T-1)$ of nodes with a waiting vehicle respectively, obtained by the ST algorithm on real instances. As the number of vehicles $M$ increases, the energy of the system increases, and vehicles have to wait more to allow other vehicles to pass. In addition, both of these quantities, after rescaled with the number of time segment $T$ or $T-1$, depend on system parameters $N$, $M$ and $T$ only via the variable $M/T (\ln N/N)^{\red{\gamma}}$, \red{with $\gamma\approx 0.95$.}

\red{The best-fit exponent of $\gamma\approx 0.95$ is obtained from the analytical results of $\avg{E}/T$ via the conventional cavity approach \req{eq_iteration}. Specifically, we obtain the analytical results of $\avg{E}/T$ as a function of $1/N$, for $M=10$ and $M=20$ as shown by the dashed line and the solid line respectively in \fig{fig_energy}. To estimate the best-fit value of $\gamma$, we first obtain the least-square polynomial fits of the dependence of $\avg{E}/T$ on $M/T (\ln N/N)^{\gamma}$ for different values of $\gamma$, for the analytical results with $M=20$. Next, the square errors between these polynomial fits and the corresponding $\gamma$-scaled analytical results of $M=10$ are computed. The value of $\gamma$ which leads to the smallest square errors gives the best-fit value of $\gamma\approx 0.95$.}

\red{To intuitively understand the scaling $M/T (\ln N/N)^{\gamma}$, we note that when the number of vehicles is small (i.e. for small $M$), the network is sparse regardless of $T$, and almost all vehicles travel via the shortest path to the destination without being blocked by other vehicles. In this case, the total cost $E$ is linearly proportional to number of vehicles $M$, and hence $E/T \propto M/T$ for the same $N$ regardless of $T$, as shown in \fig{fig_energy}. This argument holds in the regime with $E\propto M$, which a large range of values of $M$ as shown in \fig{fig_energy}, and leads to the relation $E/T \propto M/T$ . On the other hand, the quantity $\ln N/N$ is proportional to the average traffic flow per node~\cite{yeung13}, and is subject to a non-linearity with network size.}

%%%%%%%%%%%% fig 4 %%%%%%%%%%%%
\begin{figure*}
\centerline{\includegraphics[width=\linewidth]{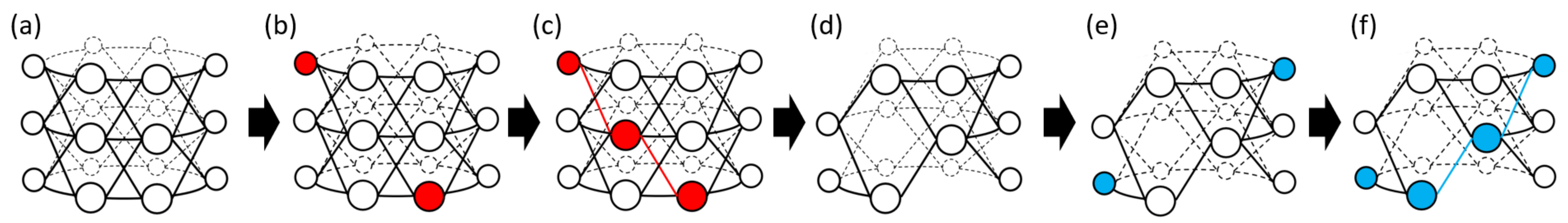}}
\caption{\red{An example of one single search by the multi-start greedy (MSG) algorithm. (a) The original space-time graph. (b) One of the vehicles is randomly picked, with its origin and destination identified on the graph. (b) The shortest path of the vehicle is identified. (c) The path of this vehicle is removed from the graph. (e) Another vehicle is randomly picked, with its origin and destination identified on the graph.  (f) The shortest path of the second vehicle on the purged graph is identified. The procedure is repeated for the next randomly drawn vehicle until either all vehicular paths are found or there is no path for some vehicles. This whole process corresponds to one single search (or one single trial) by the MSG algorithm. The algorithm is repeated for many trials, each with a random picking order of vehicles, and the lowest energy found among all the trials is denoted as $E_{\rm MSG}$.}}
\label{fig_msg}
\end{figure*}
%%%%%%%%%%%%%%%%%%%%%%%%%%

%%%%%%%%%%%% fig 4 %%%%%%%%%%%%
\begin{figure}
\vspace{-0.62cm}
%\centerline{\epsfig{figure=algorithm5.pdf, width=1\linewidth}}
\centerline{\includegraphics[width=\linewidth]{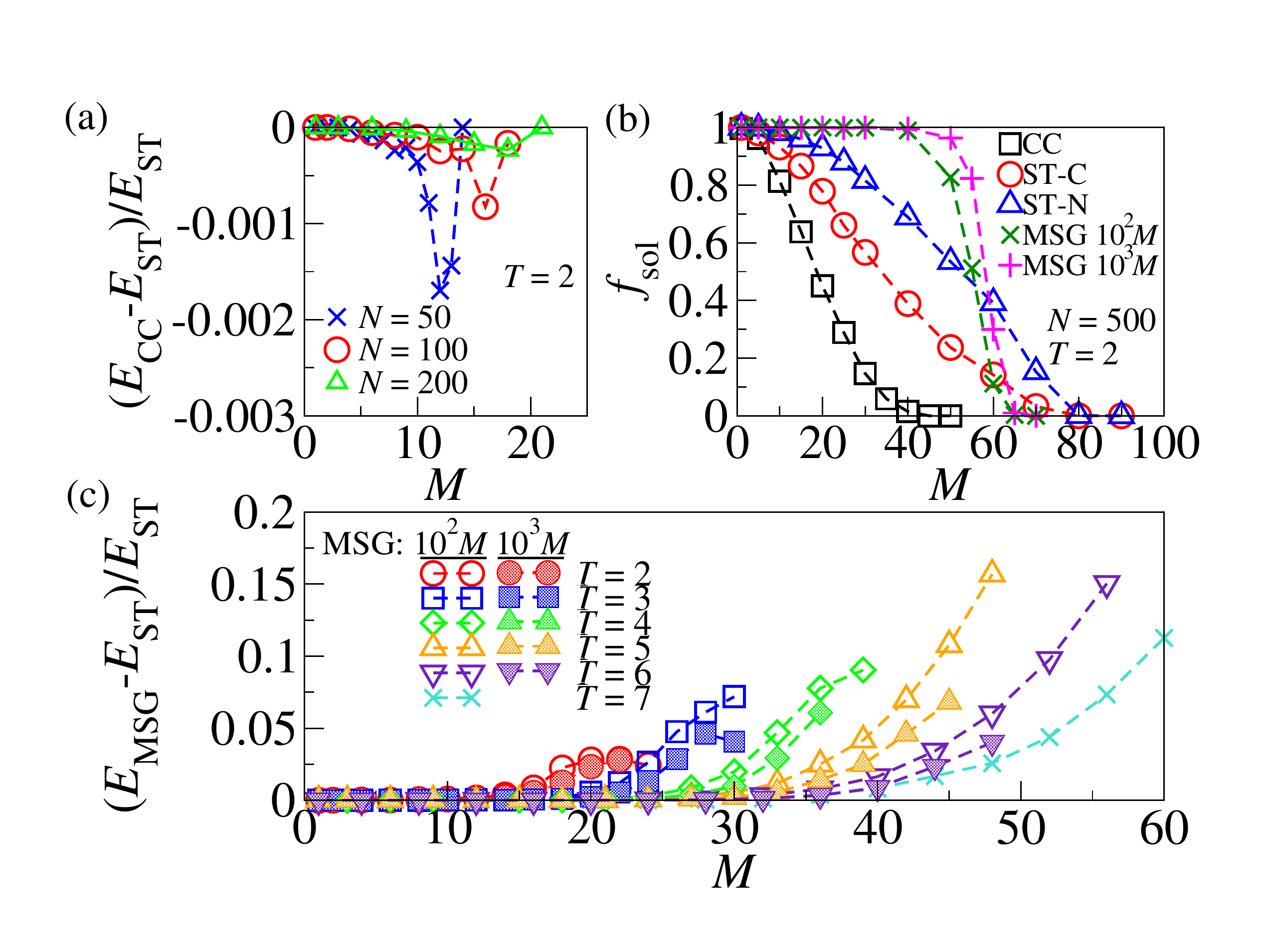}}
\caption{(a) Comparison of energy obtained by the ST and the CC algorithm. (b) Comparsion of $f_{\rm sol}$ from CC, \red{ST-C} (converged instances), \red{ST-N} (correct instances), and MSG with $10^2 M$ and $10^3 M$ starts respectively. (c) Comparison of energy obtained by the ST and MSG algorithms. \red{All results are obtained on random regular graphs with node degree $K=3$ over the converged or correct instances in at least 1000 realizations. The ST and the CC algorithms are terminated after $5\times 10^4 N$ updates if the messages do not converge. }}
\label{fig_algorithm}
\end{figure}
%%%%%%%%%%%%%%%%%%%%%%%%%%

To examine the validity of the ST algorithm\red{, one should compare its results with those obtained by the theoretical and algorithmic approaches on graphs without the structured loops on the ST graphs. As we have already seen} in \fig{fig_energy}, the \red{ST algorithmic results show} good agreement with the analytical results of conventional cavity approach \req{eq_iteration}, which is formulated without the short structured loops in ST networks. In addition, we compare the ST algorithmic results $E_{\rm ST}$ with the CC algorithmic results $E_{\rm CC}$ at $T=2$. As shown in \fig{fig_algorithm}(a), the difference is of the order $O(0.1$\%$)$. These imply that the ST algorithm (i) agrees well with both analytical and algorithmic results of the conventional cavity approach free of structured ST loops, and (ii) its validity extended to cases with $T>2$, where CC becomes computationally less feasible.

To further examine the effectiveness of the ST algorithm, we compare its results with those obtained by multi-start greedy (MSG) search~\cite{chen1996new, srinivas2003minimum}. \red{An example of a single search by the MSG algorithm is given in \fig{fig_msg}.} In each greedy search, we identify the shortest available path for a randomly drawn vehicle. The path is then removed from the ST graph and the procedure is repeated for the next randomly drawn vehicle, until either all vehicular paths are found or there is no path for some vehicles. The greedy search is repeated for either $10^2 M$ or $10^3 M$ times, \red{each with a random picking order of vehicles,} and the lowest found energy is denoted as $E_{\rm MSG}$. As shown in \fig{fig_algorithm}, ST algorithm outperforms MSG by as much as $15\%$ of cost or total travel time at large $M$ for various $T$. \red{Similar results are found in networks with degree $K>3$ as shown in Appendix~\ref{sec_degree}.}

%%%%%%%%%%%% fig 2 %%%%%%%%%%%%
\begin{figure}
%\centerline{\epsfig{figure=phase3.pdf, width=0.9\linewidth}}
\centerline{\includegraphics[width=0.9\linewidth]{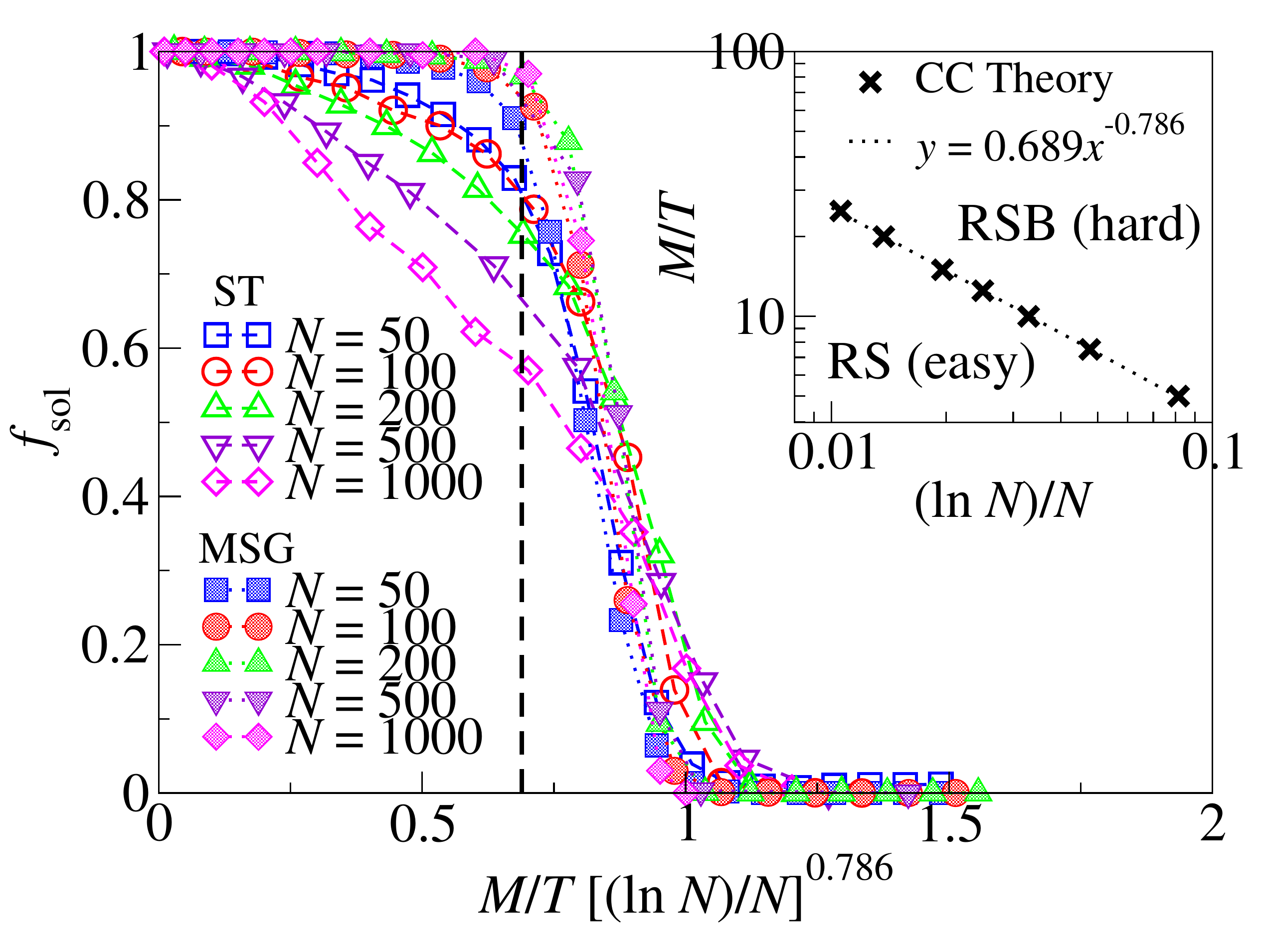}}
\caption{The fraction $f_{\rm sol}$ as a function of $M/T [(\ln N)/N]^{0.786}$, by ST and MSG. The ST algorithmic results are obtained without the bias terms in Appendix~\ref{sec_bias}. The vertical dashed line indicates the value of $M/T [(\ln N)/N]^{0.786}=0.689$, corresponding to the RS/RSB transition. Inset: theoretical results of the RS/RSB transition (see Appendix~\ref{sec_rsb}).}
\label{fig_rsb}
\end{figure}
%%%%%%%%%%%%%%%%%%%%%%%%%%

Finally, we compare $f_{\rm sol}$, i.e. the fraction of instances a solution is found by the algorithm. As shown in \fig{fig_algorithm}(c), $f_{\rm sol}$ of the ST algorithm is lower than that of MSG at intermediate values of $M$, but higher than that of MSG at large $M$, implying that ST works when MSG ceases to work. In comparison, CC has the lowest $f_{\rm sol}$ for all $M$ given the same number of iterations as the ST algorithm. To understand the behavior of $f_{\rm sol}$, we analytically identify the transition at which the replica symmetric (RS) ansatz is broken by the conventional cavity approach (see Appendix~\ref{sec_rsb}), which signals the emergence of an algorithmic-hard phase. As shown in the inset of \fig{fig_rsb}, RSB emerges when $M/T [(\ln N)/N]^{0.786}\approx 0.689$, consistent with the results of MSG and roughly consistent with those of the ST algorithm. To further improve $f_{\rm sol}$ of the ST algorithm, one can employ decimation~\cite{mezard2002random} as shown in Appendix~\ref{sec_decimation} and \fig{fig_decimation}.

%To further improve $f_{\rm sol}$ of the ST algorithm, we persistently biased vehicle paths before convergence. The $f_{\rm sol}$ of the decimated-ST algorithm and BFS are shown in \fig{fig_rsb} as a function of $M/T (N\ln N)^{0.786}$, of which the exponent is concluded from the analytical results of RS/RSB transition (see Sec. S of SI) shown in the inset of \fig{fig_rsb}. In other words, both decimated-ST and the BFS algorithms show abrupt decrease of $f_{\rm sol}$, consistent with the emergence of algorithmic-hard regime found by the conventional cavity approach.

\section{Summary}

We derived a theoretical and an algorithmic solution by conventional cavity approach for coordinating dynamical path of multiple vehicles, but they become computational intractable as the number of time segments increases. We therefore devised an alternative algorithm by applying cavity approach on the converted space-time networks. Though with the presence of structured short loops, the new algorithm show consistency with conventional cavity approach, which is free of short loops of the space-time networks. It also outperforms multi-start greedy search, and show evidence of the emergence of an algorithmic-hard regime. These results shed light on the solution to the dynamical route coordination problems as well as other dynamical problems with static analytical tools via space-time networks.

\section{Acknowledgements}

The work described in this paper was fully supported by grants from the Research Grants Council of the Hong Kong Special Administrative Region, China (Project No. EdUHK ECS 28300215, GRF 18304316, GRF 18301217).

\appendix

\section{The analytical results of $f_{\rm wait}$ by the conventional cavity approach}
\label{sec_fwait}

To compute the analytical results of $f_{\rm wait}$, i.e. the fraction of nodes with a waiting vehicle, by the conventional cavity approach, we first note that the energy $E_{\rm node, i}$ of adding a node $i$ is given by Eq. (9) without averaging over node $i$, i.e.
\begin{align}
&E_{\rm node, i}=
\\
%\left\langle
&\min_{\{\vec{\sigma}_{ii}, \{\vec{\sigma}_{ki}, \vec{\sigma}_{ik}\}|\{C_3^{i,\nu,t},C_4^{i,t}\}_{\forall \nu, t}\}}\!\left[
w_s|\vec{\sigma}_{ii}|\!+\!\sum_{k=1}^{K_i}\tE_{k}(\vec{\sigma}_{ki}, \vec{\sigma}_{ik})\right]
%\!\right\rangle.
\nonumber
\end{align}
We then compute the energy $E_{\rm node}^{\rm wait}$ of adding a node $i$ where one of the waiting links on $i$ must be occupied,
\begin{align}
\label{eq_estay}
&E_{\rm node, i}^{\rm wait}=
\\
%\left\langle\!
&\min_{\{\vec{\sigma}_{ii}, \{\vec{\sigma}_{ki}, \vec{\sigma}_{ik}\}|\{C_3^{i,\!\nu,\!t},C_4^{i,\!t},C_{\rm W}^{i}\}_{\forall \nu,\!t}\}}\!\left[
w_s|\vec{\sigma}_{ii}|\!+\!\sum_{k=1}^{K_i}\tE_{k}(\vec{\sigma}_{ki}, \vec{\sigma}_{ik})\right]
%\!\right\rangle.
\nonumber
\end{align}
where the constraint $C_{\rm W}^i$ ensures one of the waiting links on node $i$ is occupied, i.e.
\begin{align}
\label{eq_cstay}
\sum_{\nu=1}^M\left(\red{\sigma_{ii, t=0}^\nu}+\red{\sigma_{ii, t=1}^\nu}\right) = 1
\end{align}
with $\{\vec{\sigma}_{ii}, \{\vec{\sigma}_{ki}, \vec{\sigma}_{ik}\}\}$ satisfying also the constraints $C_3$ and $C_4$. Since $E_{\rm node, i}^{\rm wait}$ is computed with additional constraints compared with $E_{\rm node, i}$, hence $E_{\rm node, i}^{\rm wait}\ge E_{\rm node, i}$. The fraction $f_{\rm wait}$ of nodes with a waiting vehicle is thus given by
\begin{align}
\label{eq_fstay}
f_{\rm wait} = \Pr(E_{\rm node, i}^{\rm wait}=E_{\rm node, i})
\end{align}

Nevertheless, unlike the computation of $\avg{E_{\rm node, i}}$, the result of $f_{\rm wait}$ computed by \req{eq_fstay} is dependent on degeneracy. To further explain the complication, we define an energy $E_{\rm node}^{\rm no-wait}$ of adding a node $i$ where the waiting links on $i$ must not be occupied
\begin{align}
&E_{\rm node, i}^{\rm no-wait}=
\\
%\left\langle\!
&\min_{\substack{
\{\vec{\sigma}_{ii}, \{\vec{\sigma}_{ki}, \vec{\sigma}_{ik}\}|\{C_3^{i,\nu,t},C_4^{i,t},C_{\rm NW}^{i}\}_{\forall \nu, t}\}
}
}\!\left[
w_s|\vec{\sigma}_{ii}|\!+\!\sum_{k=1}^{K_i}\tE_{k}(\vec{\sigma}_{ki}, \vec{\sigma}_{ik})\right]
\nonumber
%\!\right\rangle.
\end{align}
where the constraint $C_{\rm NW}^i$ ensures the waiting links on node $i$ are empty, i.e.
\begin{align}
\sum_{\nu=1}^M\left(\red{\sigma_{ii, t=0}^\nu}+\red{\sigma_{ii, t=1}^\nu}\right) = 0,
\end{align}
which are the counterparts of \req{eq_estay} and \req{eq_cstay} respectively. A degeneracy in computing $E_{\rm node, i}$ would lead to 
\begin{align}
\Pr(E_{\rm node, i}^{\rm wait}=E_{\rm node, i}) + \Pr(E_{\rm node, i}^{\rm no-wait}=E_{\rm node, i}) > 1,
\end{align}
such that \req{eq_fstay} alone does not give the average ratio of waiting link occupation, but instead an upper bound.

To compare the above analytical results with the simulation results, one can define $f_{\rm wait}$ to be
\begin{align}
\label{eq_fstay2}
f_{\rm wait} = &\Pr(E_{\rm node, i}^{\rm wait}=E_{\rm node, i}, E_{\rm node, i}^{\rm no-wait}\neq E_{\rm node, i}) 
\nonumber\\
&+ \frac{1}{2} \Pr(E_{\rm node, i}^{\rm wait}=E_{\rm node, i}, E_{\rm node, i}^{\rm no-wait}= E_{\rm node, i}),
\end{align}
which corresponds to only an estimate of $f_{\rm wait}$ since the factor $\frac{1}{2}$ is arbitrary in the above equation. Alternatively, one can introduce a small random bias term $\epsilon_{ij,t}^\nu$ for the directed link $ij$ and vehicle $\nu$ at time $t$ in \req{eq_iteration}, in the recursion of $E_{ij}(\vec{\sigma}_{ij}, \vec{\sigma}_{ji})$, given by
\begin{widetext}
\begin{align}
\label{eq_iterationDelta}
&E_{i\to j}(\vec{\sigma}_{ij}, \vec{\sigma}_{ji}) 
=
\nonumber\\
& \min_{\{\vec{\sigma}_{ii}, \{\vec{\sigma}_{ki}, \vec{\sigma}_{ik}\}|\{C_3^{i, \nu, t}, C_4^{i, t}, C_5^{ik, t}\}_{\forall k, \nu, t}\}}\left[|\vec{\sigma}_{ij}|
+|\vec{\sigma}_{ji}|+w_s|\vec{\sigma}_{ii}|+\sum_{k}a_{ki}E_{k\to i}(\vec{\sigma}_{ki}, \vec{\sigma}_{ik})\right]
%\nonumber\\
%&
+\sum_{\nu, t}(\sigma_{ij, t}^\nu\epsilon_{ij,t}^\nu + \sigma_{ji, t}^\nu\epsilon_{ji,t}^\nu),
\end{align}
\end{widetext}
which breaks the degeneracy of the computation of $E_{\rm node, i}$. We remark that the purpose of the random bias term $\epsilon_{ij,t}^\nu$ is only to break the degeneracy, but not to interfere the identification of paths without degeneracy. In this case, the magnitude of $\epsilon_{ij,t}^\nu$ should satisfy $|\epsilon_{ij,t}^\nu| \ll 1/T_{\rm iteration}$, where $T_{\rm iteration}$ is the maximum number of iteration in the population dynamics for solving \req{eq_iterationDelta}. Hence, \req{eq_fstay} can be used to compute the correct $f_{\rm wait}$. We found that the results of $f_{\rm wait}$ computed through \req{eq_iterationDelta} and that computed by \req{eq_fstay2} are similar.

The results of $f_{\rm wait}$ are shown in the inset of \fig{fig_energy}. We scale $f_{\rm wait}$ with $T-1$ instead of $T$ and show the results of $f_{\rm wait}/(T-1)$ because at most $T-1$ waiting links of a node can be occupied, as no vehicle would just stay the whole period of $T$ without traveling.

\section{The simplfied recursion relations on space-time network}
\label{sec_simSP}

Equations~(\ref{eq_forward}) and (\ref{eq_backward}) can be used as an algorithm to coordinate multi-vehicle dynamical routes on networks, we can simplify these equations by implementing the constraints $C_3, C_4$ and $C_5$. We first denote
\begin{align}
\esp_{\ijtf, \nu} = \esp_{\ijtf}(\vec{\sigma}_{\ijt}=\vec{v}_\nu)
\\
\esp_{\ijtf, 0} = \esp_{\ijtf}(\vec{\sigma}_{\ijt}=\vec{0})
\end{align}
where $\vec{x}_\nu$ is a unit-vector with the $\nu$-th entry to be 1 and all the other entries to be 0 ; $\vec{0}$ is a zero vector. We then define
\begin{align}
\label{eq_weight}
w_{ij} = 1+\delta_{ij}(w_s-1)
\end{align}
to be the cost (or energy) on the link $ij$. One can then simplify the forward message recursion \req{eq_forward} as
\begin{widetext}
\begin{align}
\label{eq_forwardSimMu}
\esp_{\ijtf, \nu} = \sum_k a_{ki}\esp_{\kitf, 0} + \sum_{k\neq j} a_{ki}\esp_{\kitbb, 0}
%\nonumber\\
%&
+
\begin{cases}
\displaystyle
w_{ij}\!+\!\min_{\{k|a_{ki}=1\}}\left[\esp_{\kitf, \nu}-\esp_{\kitf, 0}\right],  
&\mbox{if $\Lambda^\nu_{i,t}=0, \forall\nu$,}
\\
\displaystyle
w_{ij}, 
&\mbox{if $\Lambda^\nu_{i,t} = 1$,}
\\
\infty, 
&\mbox{if $\Lambda^\nu_{i,t} = -1$,}
\\
\infty, 
&\mbox{if $\Lambda^\mu_{i,t}\!=\!1$, $\exists\mu\neq\nu$,}
\\
\displaystyle
w_{ij}\!+\!\min_{\{k|a_{ki}=1\}}\left[\esp_{\kitf, \nu}-\esp_{\kitf, 0}\right],  
&\mbox{if $\Lambda^\mu_{i,t}\!=\!-1, \exists \mu\!\neq\!\nu$.}
\end{cases}
\end{align}
\begin{align}
\label{eq_forwardSimZero}
\esp_{\ijtf, 0} &= \sum_k a_{ki}\esp_{\kitf, 0} + \sum_{k\neq j} a_{ki}\esp_{\kitbb, 0}
\nonumber\\
&+
\begin{cases}
\displaystyle
\min\left\{0, \min_{\{k, l, \nu|l\neq j, a_{ki}a_{li}=1\}}\left[\esp_{\kitf, \nu}-\esp_{\kitf, 0} + \esp_{\litb, \nu}-\esp_{\litb, 0}\right]\right\},
%\\
%&\hspace{-3cm}
&\mbox{if $\Lambda^\nu_{i,t} = 0, \forall\nu$,}
\\
\displaystyle
\min_{\{k|k\neq j, a_{ki}=1\}}\left[\esp_{\kitbb, \nu}-\esp_{\kitbb, 0}\right], 
%&\hspace{-3cm}
&\mbox{if $\Lambda^\nu_{i,t} = 1$,}
\\
\displaystyle
\min\Bigg\{0, \min_{\{k, \nu|a_{ki}=1, \Lambda^\nu_{i,t}=-1\}}\left[\esp_{\kitf, \nu}-\esp_{\kitbb, 0}\right],
\\
\hspace{0.6cm}\min_{\{k, l, \nu|l\neq j, a_{ki}a_{li}\!=\!1, \Lambda^\nu_{i,t}=0\}}\left[\esp_{\kitf, \nu}-\esp_{\kitf, 0} + \esp_{\litb, \nu}-\esp_{\litb, 0}\right]\Bigg\},
%\\
%&\hspace{-3cm}
&\mbox{if $\Lambda^\nu_{i,t}\!=\!-1, \exists\nu$.}
\end{cases}
\end{align}
\end{widetext}
where $\esp_{\ijtf, \nu}=\infty$ corresponds to cases where not all the constraints are satisfied. For simplicity, we do not consider scenarios where a node can be the origin of one vehicle and the destination of another vehicle. The equations of this scenario can be similarly derived but with more complication.

Since $\esp$ is an extensive quantity, one can define an intensive quantity $\etsp$ according to Eqs. (\ref{eq_etspf}) and (\ref{eq_etspb}). We then subtract \req{eq_forwardSimMu} by \req{eq_forwardSimZero}, and with the definition of $\etsp$, one can further simplify the forward message recursion \req{eq_forward} as
\begin{widetext}
\begin{align}
\label{eq_forwardFinal}
&\etsp_{\ijtf, \nu} = 
\nonumber\\
&\quad
\begin{cases}
\displaystyle
w_{ij} + \min_{\{k|a_{ki}=1\}}\left[\etsp_{\kitf, \nu}\right] 
%\\
%\displaystyle
%\hspace{1cm}
- \min\left\{0, \min_{\{k, l, \nu|l\neq j, \mu\neq\nu, a_{ki}a_{li}=1\}}\left[\etsp_{\kitf, \mu}+ \etsp_{\litb, \mu}\right]\right\}, 
%\\
&\hspace{-2cm}
\mbox{if $\Lambda^\nu_{i,t}\!=\!0, \forall\nu$,} 
\\
\displaystyle
w_{ij} - \min_{\{k|k\neq j, a_{ki}=1\}}\left[\etsp_{\kitbb, \nu}\right],
&\hspace{-2cm}
\mbox{if $\Lambda^\nu_{i,t} = 1$,}
\\
\infty, 
&\hspace{-2cm}
\mbox{if $\Lambda^\nu_{i,t} = -1$,}
\\
\infty, 
&\hspace{-2cm}
\mbox{if $\Lambda^\mu_{i,t} = 1, \exists\mu\neq\nu$,}
\\
\displaystyle
w_{ij} + \min_{\{k|a_{ki}=1\}}\left[\etsp_{\kitf, \nu}\right] 
\\
\displaystyle
\quad\quad- \min\left\{0, 
\min_{\{k, \mu|a_{ki}=1, \Lambda^\mu_{i,t}=-1, \mu\neq\nu\}}\left[\etsp_{\kitf, \mu}\right],
\min_{\{k, l, \mu|l\neq j, \mu\neq\nu, a_{ki}a_{li}=1\}}\left[\etsp_{\kitf, \mu}+ \etsp_{\litb, \mu}\right]\right\}, 
\\
&\hspace{-2cm}
\mbox{if $\Lambda^\mu_{i,t}=-1, \exists \mu\neq\nu$.}
\end{cases}
\end{align}

Similarly, the backward message recursion relation \req{eq_backward} can be simplified and becomes 
\begin{align}
\label{eq_backwardFinal}
&\etsp_{\ijtb, \nu} = 
\nonumber\\
&\quad
\begin{cases}
\displaystyle
w_{ij} + \min_{\{k|a_{ki}=1\}}\left[\etsp_{\kitbb, \nu}\right]
%\\
%\displaystyle
%\hspace{1cm}
- \min\left\{0, \min_{\{k, l, \mu|k\neq j, \mu\neq\nu, a_{ki}a_{li}=1\}}\left[\etsp_{\kitf, \mu}+ \etsp_{\litb, \mu}\right]\right\}, 
&\hspace{-2.5cm}
\mbox{if $\Lambda^\nu_{i,t}\!=\!0, \forall\nu$,} 
\\
\displaystyle
\infty,
&\hspace{-2.5cm}
\mbox{if $\Lambda^\nu_{i,t} = 1$,}
\\
\displaystyle
w_{ij} 
%\\
%\displaystyle
- \min\left\{0, 
\min_{\{k|k\neq j, a_{ki}=1, \Lambda^\nu_{i,t}=-1\}}\left[\etsp_{\kitf, \nu}\right],
\min_{\{k, l, \mu|k\neq j, \mu\neq\nu, a_{ki}a_{li}=1\}}\left[\etsp_{\kitf, \mu}+ \etsp_{\litb, \mu}\right]\right\}, 
\\
&\hspace{-2.5cm}
\mbox{if $\Lambda^\nu_{i,t} = -1$,}
\\
\infty, 
&\hspace{-2.5cm}
\mbox{if $\Lambda^\mu_{i,t} = 1, \exists\mu\neq\nu$,}
\\
\displaystyle
w_{ij} + \min_{\{k|a_{ki}=1\}}\left[\etsp_{\kitbb, \nu}\right] 
\\
\displaystyle
\quad\quad- \min\left\{0, 
\min_{\{k, \mu|k\neq j, a_{ki}=1, \Lambda^\mu_{i,t}=-1, \mu\neq\nu\}}\left[\etsp_{\kitf, \mu}\right],
\min_{\{k, l, \mu|k\neq j, \mu\neq\nu, a_{ki}a_{li}=1\}}\left[\etsp_{\kitf, \mu}+ \etsp_{\litb, \mu}\right]\right\}, 
\\
&\hspace{-2.5cm}
\mbox{if $\Lambda^\mu_{i,t}=-1, \exists \mu\neq\nu$.}
\end{cases}
\end{align}
We remark that $a_{ii}=1$ for all $i$ in the space-time network, and $\etsp_{\ijtf, 0} = \etsp_{\ijtb, 0} = 0$ by the definition Eqs.~(\ref{eq_etspf}) and (\ref{eq_etspb}). Equations (\ref{eq_forwardFinal}) and (\ref{eq_backwardFinal}) thus constitute a message-passing algorithm on the space-time network. After the messages converge on all links, the optimal configuration can be found by \req{eq_config}, which can be simplified as the following equations
\begin{align}
\label{eq_configSim}
\vec{\sigma}_{\ijt}^*
=
\begin{cases}
\vec{x}_{\nu^*}, & 
\mbox{if $\etsp_{\ijtf, \nu} + \etsp_{\ijtf, \nu} - w_{ij} < 0$,
such that $\nu^*=\argmin\left\{\etsp_{\ijtf, \nu}\!+\!\etsp_{\ijtf, \nu}\!-\![1\!+\!\delta_{ij}(w_s\!-\!1)]\right\}$}
%\\
%& 
%\mbox{such that $\nu^*=\argmin\left\{\etsp_{\ijtf, \nu} + \etsp_{\ijtf, \nu} - [1\!+\!\delta_{ij}(w_s\!-\!1)]\right\}$}
%\\
\\
\vec{0}, & \mbox{if $\etsp_{\ijtf, \nu} + \etsp_{\ijtf, \nu} - w_{ij} \ge 0 $}
\end{cases}
\end{align}
\end{widetext}

In summary, Eqs.~(\ref{eq_forwardFinal}) -- (\ref{eq_configSim}) constitute an algorithm which can be used to optimize and coordinate the dynamical routes of multiple traveling vehicles between their respective origins and destinations. 

\section{Identification of solution in cases of degeneracy}
\label{sec_bias}

Nevertheless, Eqs.~(\ref{eq_forwardFinal}) -- (\ref{eq_configSim}) may result in a solution of multiple degenerate paths for a single vehicle, and does not identify specifically one route for each vehicle. For the algorithm to select a single route for each vehicle, we introduce a random bias term $\epsilon_{ij}$ in the cost of each link, given by 
\begin{align}
\label{eq_weightBiase}
w_{ij} = 1+\delta_{ij}(w_s-1)+\epsilon_{ij},
\end{align}
with $\epsilon_{ij} = \epsilon_{ji}$. We then use \req{eq_weightBiase} instead of \req{eq_weight} in Eqs.~(\ref{eq_forwardFinal}) -- (\ref{eq_configSim}) as an algorithm to optimize real instances.

Since the purpose of the term $\epsilon_{ij}$ is to break degeneracy, but not to interfere the identification of paths without degeneracy. In this case, the magnitude of $\epsilon_{ij,t}^\nu$ should satisfy $|\epsilon_{ij,t}^\nu| < 1/N$. Based on our results, we also found that the convergence time is greatly shortened if $\epsilon_{ij}$ is a multiple of a small rational number much less than 1, e.g. 0.001.

\section{The dependence on network degree}
\label{sec_degree}

%%%%%%%%%%%% fig 2 %%%%%%%%%%%%
\begin{figure}
\centerline{\epsfig{figure=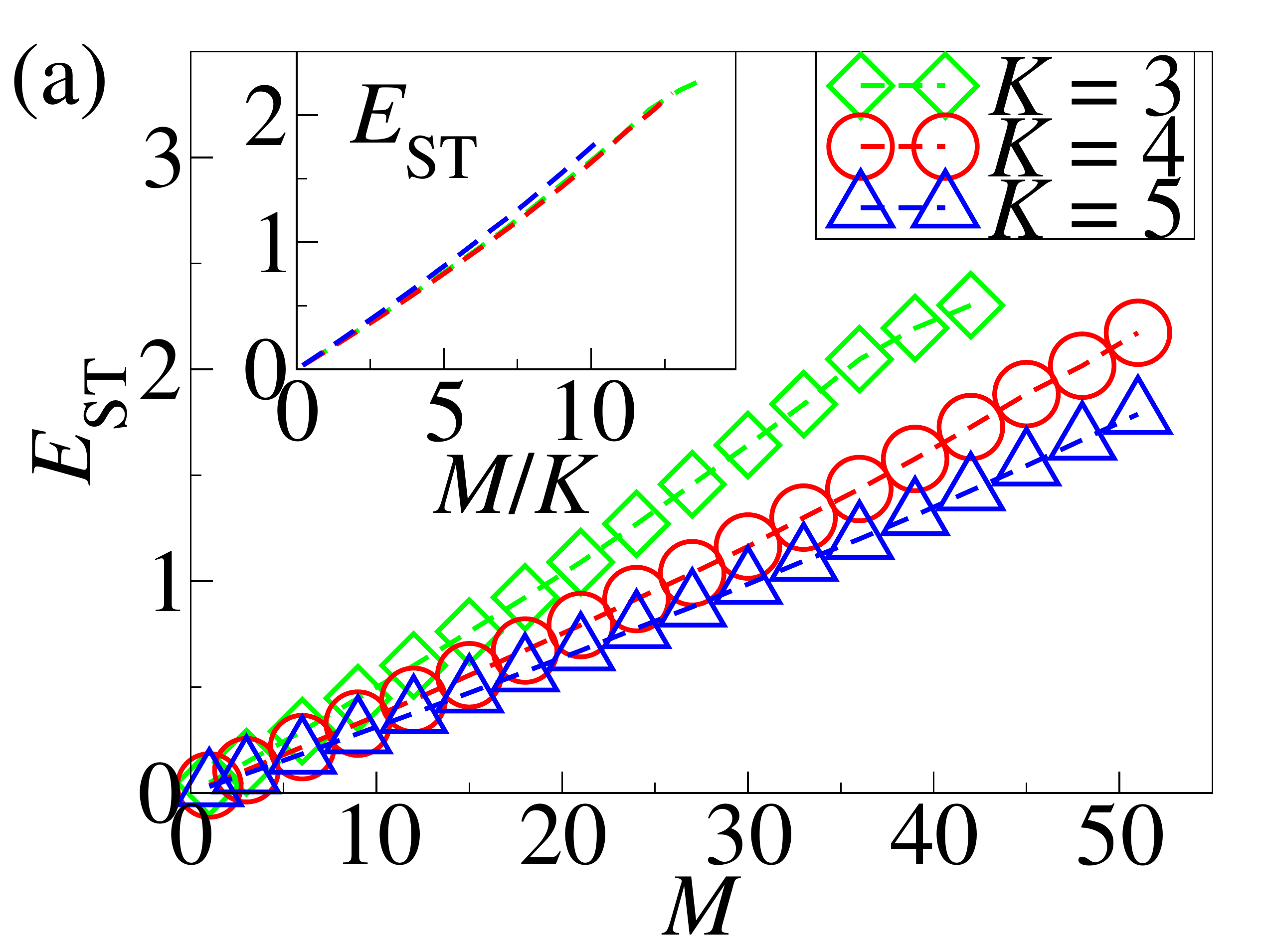, width=0.5\linewidth}
\epsfig{figure=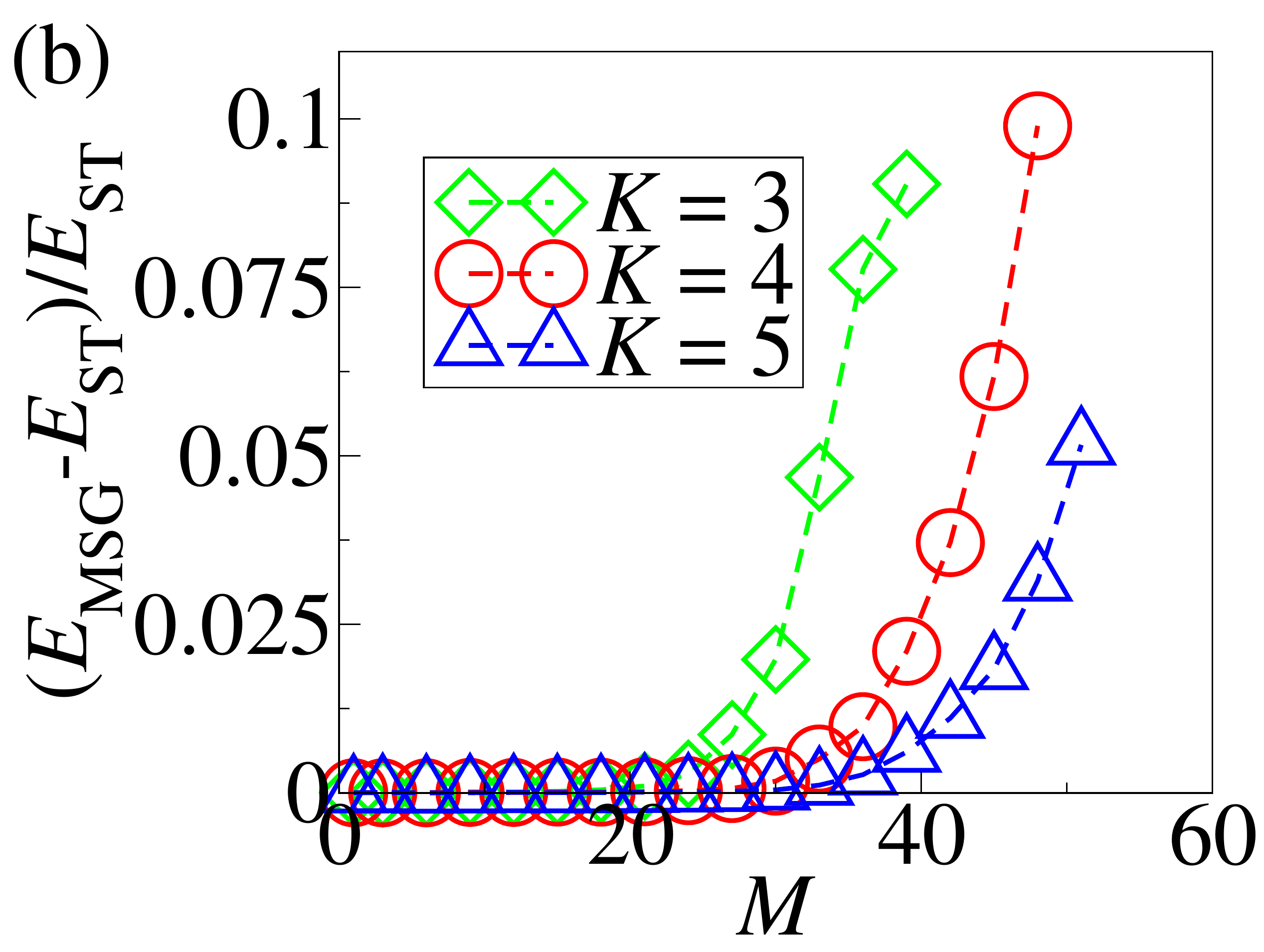, width=0.5\linewidth}}
\caption{\red{(a) The average energy $E_{\rm ST}$ of the optimal path configurations as a function of the number of vehicles $M$, obtained by the ST algorithm on random regular graphs with $N=100$, $T=4$ and $K=3,4,5$, averaged over the converged instances in at least 1000 realizations. The algorithm is terminated after $5\times 10^4 N$ updates if the messages do not converge. Inset: the average energy $E_{\rm ST}$ as a function of $M/K$. The symbols on the line are omitted for the sake of clarity in illustration. (b) Comparison of the energy obtained by the ST algorithm and the MSG algorithm on random regular graphs with $N=100$, $T=4$ and $K=3,4$ and $5$.}}
\label{fig_K}
\end{figure}
%%%%%%%%%%%%%%%%%%%%%%%%%%

\red{As we have mentioned in Sec.~\ref{sec_analytical} and \ref{sec_algorithm}, the optimization algorithm derived from the conventional cavity approach has a computational complexity of $O[(2M+1)^{T(K-1)}]$, which greatly increases with node degree $K$ even with an intermediate value of $M$. On the other hand, the ST algorithm has a computational complexity of $O(MKT)$ as discussed in Sec.~\ref{sec_algorithm}, which is computational feasible even for networks with large $K$.}

\red{To examine the dependence of the optimized path configuration on node degree, we implement the ST algorithm on random regular graphs with $K>3$. The results of the optimized average energy $E_{\rm ST}$ are shown in \fig{fig_K}(a). As we can see, the energy of the networks with the same number of vehicles (i.e. the same value of $M$) decreases with increasing $K$. It is because the larger the node degree, the more the links in the network, and the higher the capacity of the network to accommodate traffic flows. In this case, vehicles are less often blocked by other vehicles on networks with large $K$, leading to a smaller energy. We therefore show the optimized energy $E_{\rm ST}$ as a function of $M/K$ in the inset of \fig{fig_K}(a). The collapse of the results suggests that the optimized energy is roughly proportional to $M/K$, and hence the capacity of the network is proportional to $K$.}

\red{We further compare the results of the ST algorithm with those of the MSG algorithm on networks with various node degrees. As shown in \fig{fig_K}(b), the ST algorithm outperforms the MSG algorithm in cases with various values of $K$, similar to the results for $K=3$ as shown in \fig{fig_algorithm}(c).}

\section{The RS/RSB transition in the dynamical route coordination problem}
\label{sec_rsb}

%%%%%%%%%%%% fig 2 %%%%%%%%%%%%
\begin{figure}
\centerline{\epsfig{figure=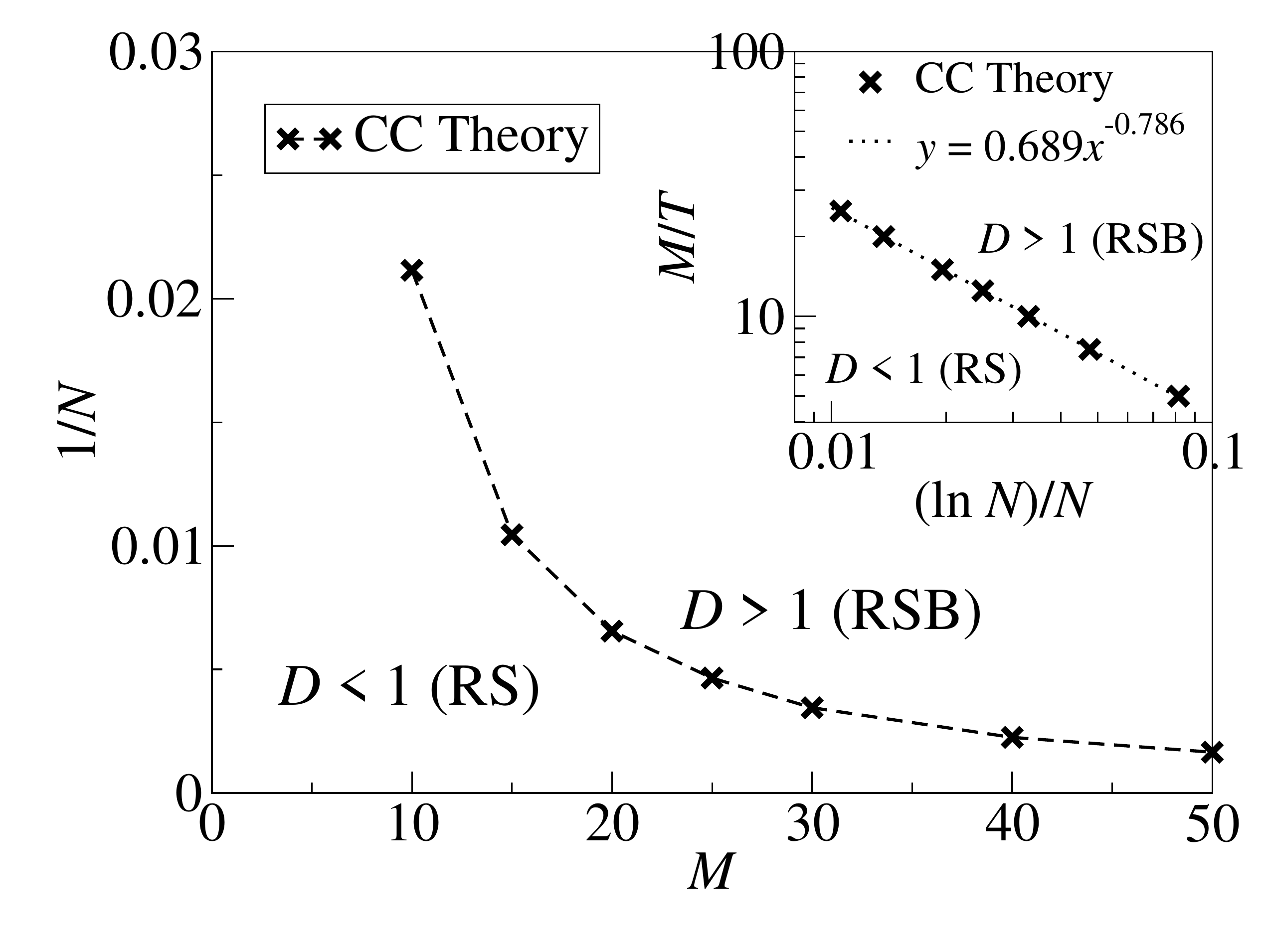, width=0.9\linewidth}}
\caption{The analytical results of the emergence of the replica-symmetry broken (RSB) phase (i.e. the phase with $D>1$), obtained by solving \req{eq_rsb} for $T=2$ by population dynamics with a pool size of 50000 and $2.5\times 10^8$ updates. Inset: the same results in log-log plot with $M/T$ as the vertical axis and $(\ln N)/N$ as the horizontal axis; the dotted line is the best-fit line.}
\label{fig_rsb}
\end{figure}
%%%%%%%%%%%%%%%%%%%%%%%%%%

The transition from an algorithmic-easy phase to an algorithmic-hard phase in hard optimization problems is \red{originally found to be related} to the transition from the so-called replica-symmetry (RS) phase to the replica-symmetry broken (RSB) phase in the studies of spin glass~\cite{mezard87, nishimori01}. \red{In later studies, various transitions are identified in the parameter regime before the emergence of RS/RSB transition~\cite{zdeborava07}. Observations in some combinatorial optimization problems have suggested that the transition between the algorithmic-easy and -hard phase coincides with the \emph{rigidity} or \emph{freezing transition}, beyond which all solution clusters in the solution space contain \emph{frozen variables} (i.e. variables which take the same value in all solutions of the solution cluster)~\cite{zdeborava07}}. Here, we follow the approach in~\cite{yeung13} to identify the phase transition between the RS and the RSB phase \red{and examine its relation with the emergence of algorithmic-hard phase} in the dynamical route coordination problem at $T=2$. 

We first write down the following recursion of the joint functional probability distribution $P[E^\alpha_{i\to j}(\vec{\sigma}_{ij}, \vec{\sigma}_{ji}), E^\beta_{i\to j}(\vec{\sigma}_{ij}, \vec{\sigma}_{ji})]$ of two energy functions $E^\alpha_{i\to j}(\vec{\sigma}_{ij}, \vec{\sigma}_{ji})$ and $E^\beta_{i\to j}(\vec{\sigma}_{ij}, \vec{\sigma}_{ji})$ labeled by $\alpha$ and $\beta$, given by
\begin{widetext}
\begin{align}
\label{eq_rsb}
&P[E^\alpha_{i\to j}(\vec{\sigma}_{ij}, \vec{\sigma}_{ji}), E^\beta_{i\to j}(\vec{\sigma}_{ij}, \vec{\sigma}_{ji})]
\nonumber\\
&=
\sum_{K=1}^\infty \frac{K P(K)}{\avg{K}}
\int d\vec{\Lambda} P(\vec{\Lambda})
\prod_{k=1}^{K-1}\int 
dE^\alpha_{k\to i}(\vec{\sigma}_{ki}, \vec{\sigma}_{ik})
dE^\beta_{k\to i}(\vec{\sigma}_{ki}, \vec{\sigma}_{ik})
P[E^\alpha_{k\to i}(\vec{\sigma}_{ki}, \vec{\sigma}_{ik}), E^\beta_{k\to i}(\vec{\sigma}_{ki}, \vec{\sigma}_{ik})]
\nonumber\\
&\times\delta\left\{\!E^\alpha_{i\to j}(\vec{\sigma}_{ij}, \vec{\sigma}_{ji})
\!-\!\min_{\{\vec{\sigma}_{ii}, \{\vec{\sigma}_{ki}, \vec{\sigma}_{ik}\}|\{C_3^{i, \nu, t}, C_4^{i, t}, C_5^{ik, t}\}_{\forall k, \nu, t}\}}
\!\left[\!|\vec{\sigma}_{ij}|
\!+\!|\vec{\sigma}_{ji}|\!+\!w_s|\vec{\sigma}_{ii}|\!+\!\sum_{k=1}^{K-1}E^\alpha_{k\to i}(\vec{\sigma}_{ki}, \vec{\sigma}_{ik})\!\right]\!\right\}
\nonumber\\
&\times\delta\left\{\!E^\beta_{i\to j}(\vec{\sigma}_{ij}, \vec{\sigma}_{ji})
\!-\!\min_{\{\vec{\sigma}_{ii}, \{\vec{\sigma}_{ki}, \vec{\sigma}_{ik}\}|\{C_3^{i, \nu, t}, C_4^{i, t}, C_5^{ik, t}\}_{\forall k, \nu, t}\}}
\!\left[\!|\vec{\sigma}_{ij}|
\!+\!|\vec{\sigma}_{ji}|\!+\!w_s|\vec{\sigma}_{ii}|\!+\!\sum_{k=1}^{K-1}E^\beta_{k\to i}(\vec{\sigma}_{ki}, \vec{\sigma}_{ik})\!\right]\!\right\}
\end{align}
\end{widetext}
where $P(K)$ and $P(\Lambda)$ are the degree distribution and the distribution of origin and destination, respectively. In this case, the recursion of the two energy functions $\alpha$ and $\beta$ always follow the same quenched disorders during the iteration\red{, and are two replicas of the same system}.

To identify the RSB transition, we start with an initial condition of $P[E^\alpha_{i\to j}, E^\beta_{i\to j}]$ with non-zero contribution in the domain of $E^\alpha_{i\to j} \neq E^\beta_{i\to j}$, \red{and examine if \req{eq_rsb} converges to a distribution $P[E^\alpha_{i\to j}, E^\beta_{i\to j}]$ with non-zero domain only along the diagonal of $E^\alpha_{i\to j} = E^\beta_{i\to j}$, i.e. converges to the same state from different initial conditions. Nevertheless, since it is difficult to compute a complete solution of $P[E^\alpha_{i\to j}, E^\beta_{i\to j}]$ owing to its high-dimensional functional domain, we therefore iterate \req{eq_rsb}} with different initial condition of $E^\alpha_{i\to j}$ and  $E^\beta_{i\to j}$ but the same quenched disorders\red{, and examine the change of hamming distance between the two replicas $\alpha$ and $\beta$ at consecutive recursion layers}. We then measure the \red{ratio $D$} given by
\begin{align}
&D = 
\nonumber\\
&\frac{\sqrt{\sum_{ \vec{\sigma}_{ij}, \vec{\sigma}_{ji} }
\!\left[E^\alpha_{i\to j}(\vec{\sigma}_{ij}, \vec{\sigma}_{ji})\!-\!E^\beta_{i\to j}(\vec{\sigma}_{ij}, \vec{\sigma}_{ji})\right]^2} 
}{
\frac{1}{K\!-\!1}\sum_{k\!=\!1}^{K\!-\!1}\sqrt{\sum_{ \vec{\sigma}_{ki}, \vec{\sigma}_{ik} }
\!\left[E^\alpha_{k\to i}(\vec{\sigma}_{ki}, \vec{\sigma}_{ik})\!-\!E^\beta_{k\to i}(\vec{\sigma}_{ki}, \vec{\sigma}_{ik})\right]^2} },
\end{align}
\red{where the numerator denotes the hamming distance between the two replicas at node $i$, and the denominator denotes the average hamming distance among the descendants $k$ of node $i$.}
If the \red{ratio} $D\le 1$, \red{the hamming distance between the two replicas decreases as the recursion proceeds. In this case, the two replicas with different boundary conditions would eventually converge to the same state, and }the system is replica symmetric (RS). On the other hand, if $D>1$, \red{the hamming distance of the two replicas increases as the recursion proceeds. In this case, even a small initial difference in the boundary condition of the two replicas would be enlarged, resulting in two different states; }the system is in the replica symmetry broken (RSB) phase. The analytical results of the RS/RSB transition identified by the value of $D$ is shown in \fig{fig_rsb}, where the RSB phase emerges when $M/T[(\ln N)/N]^{-0.786} > 0.689$.

\section{Decimation in the ST algorithm}
\label{sec_decimation}

%%%%%%%%%%%% fig 2 %%%%%%%%%%%%
\begin{figure}[h]
\centerline{\epsfig{figure=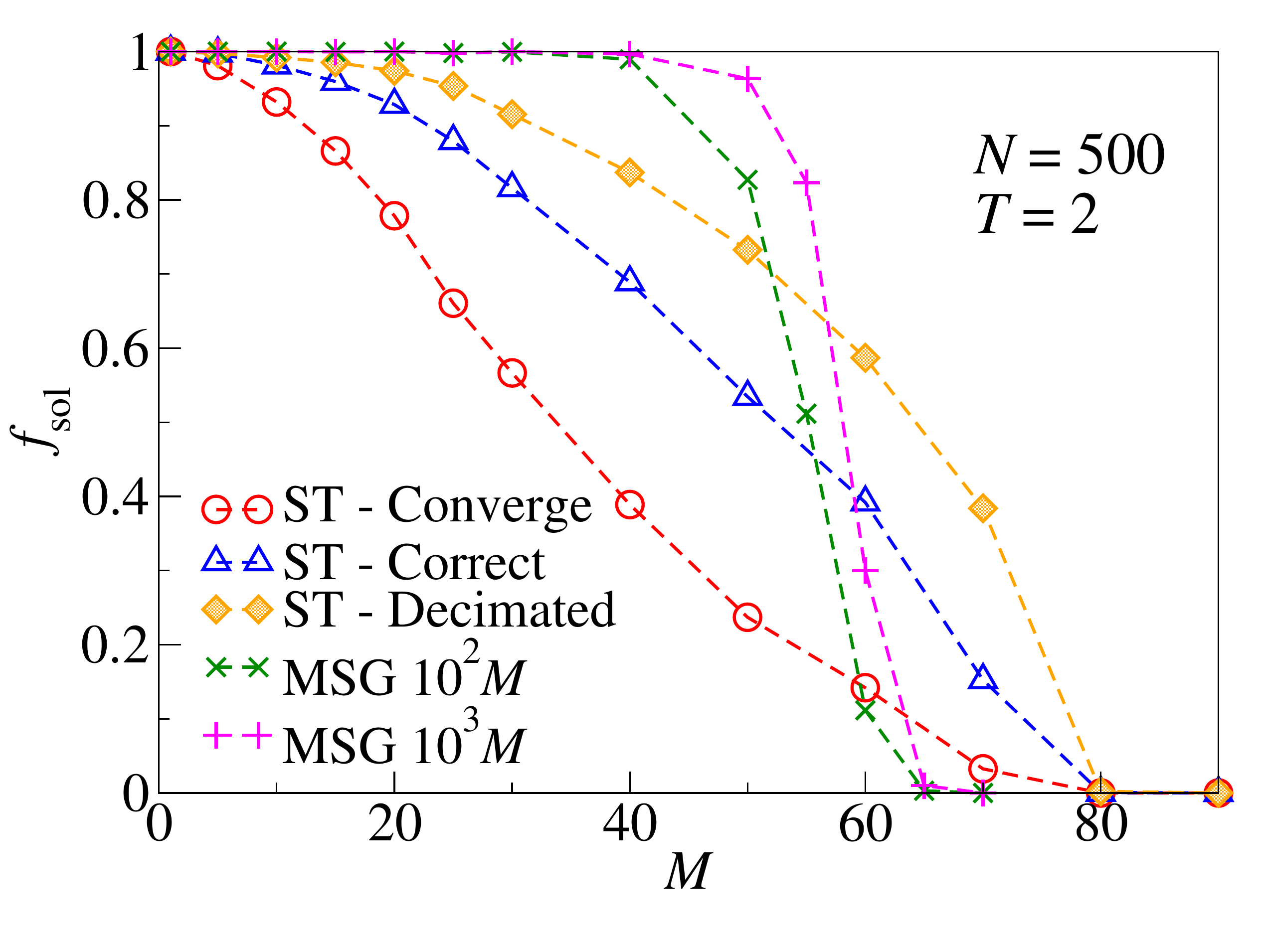, width=0.9\linewidth}}
\caption{Comparison of $f_{\rm sol}$ obtained from ST-Converge (converged instances), ST-Correct (correct instances), ST-Decimated, and multi-start greedy (MSG) search with $10^2 M$ and $10^3 M$ starts respectively.}
\label{fig_decimation}
\end{figure}
%%%%%%%%%%%%%%%%%%%%%%%%%%

To further improve $f_{\rm sol}$ of the ST algorithm, we employ decimation as in other optimization problems~\cite{mezard2002random} and fix the path of vehicle where the un-converged messages indicate a path persistently. We show the results of the decimated-ST algorithm in \fig{fig_decimation}, compared to the results we show in \fig{fig_algorithm}(b). As we can see, $f_{\rm sol}$ obtained by the decimated-ST algorithm is higher than that obtained from the multi-start greedy (MSG) search with $10^2 M$ and $10^3 M$ starts \red{at the large values of $M\gtrsim 60$}, as well as the fraction of the converged or the correct instances in the ST algorithm \red{for all values of $M$}. These results show that the ST algorithm is able to obtain a solution when MSG ceases to work.

\section{Algorithmic results on graphs with multiple lanes}
\label{sec_lane}

%%%%%%%%%%%% fig 2 %%%%%%%%%%%%
\begin{figure}[!t]
\centerline{
\raisebox{0.3cm}{\epsfig{figure=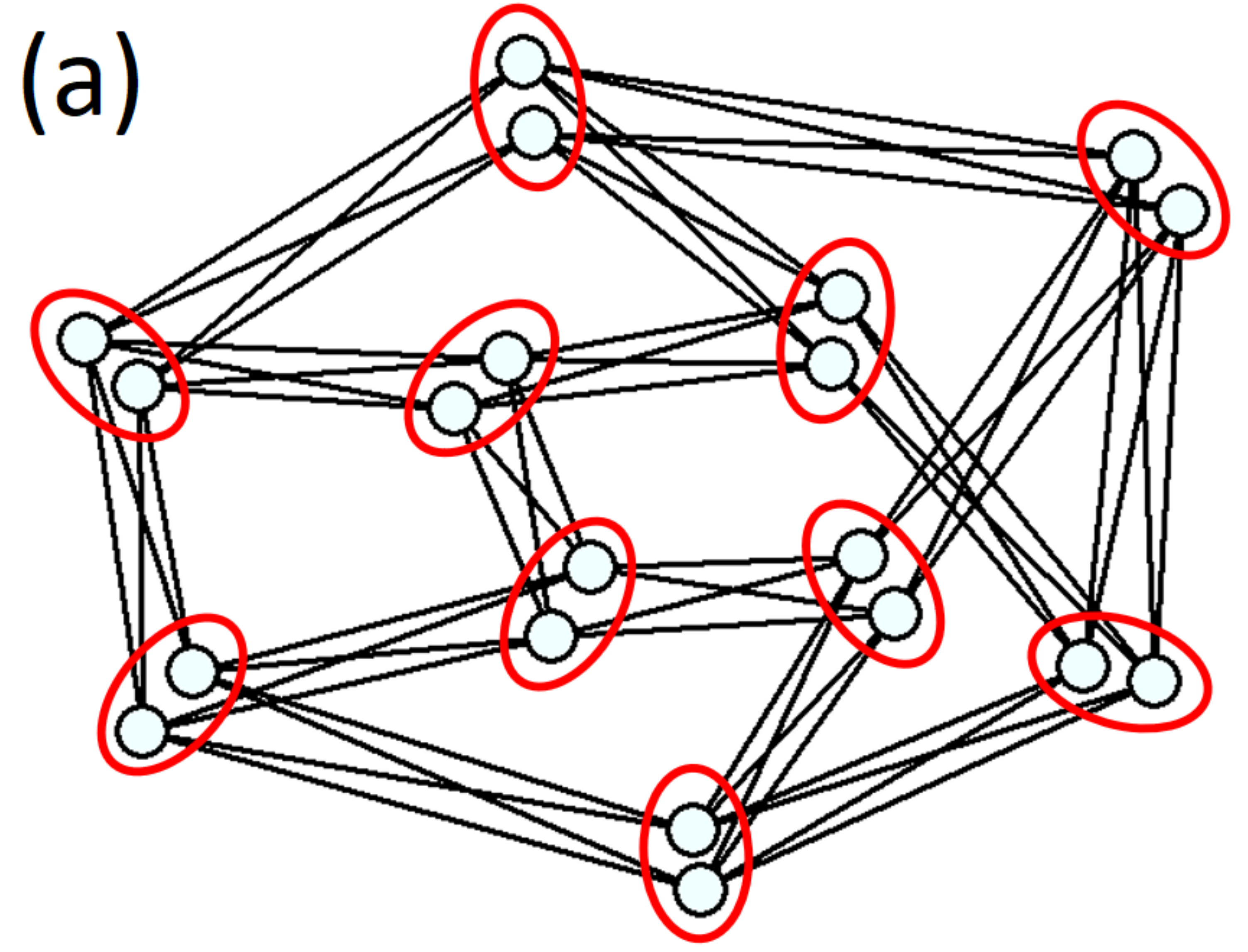, width=0.45\linewidth}}
\epsfig{figure=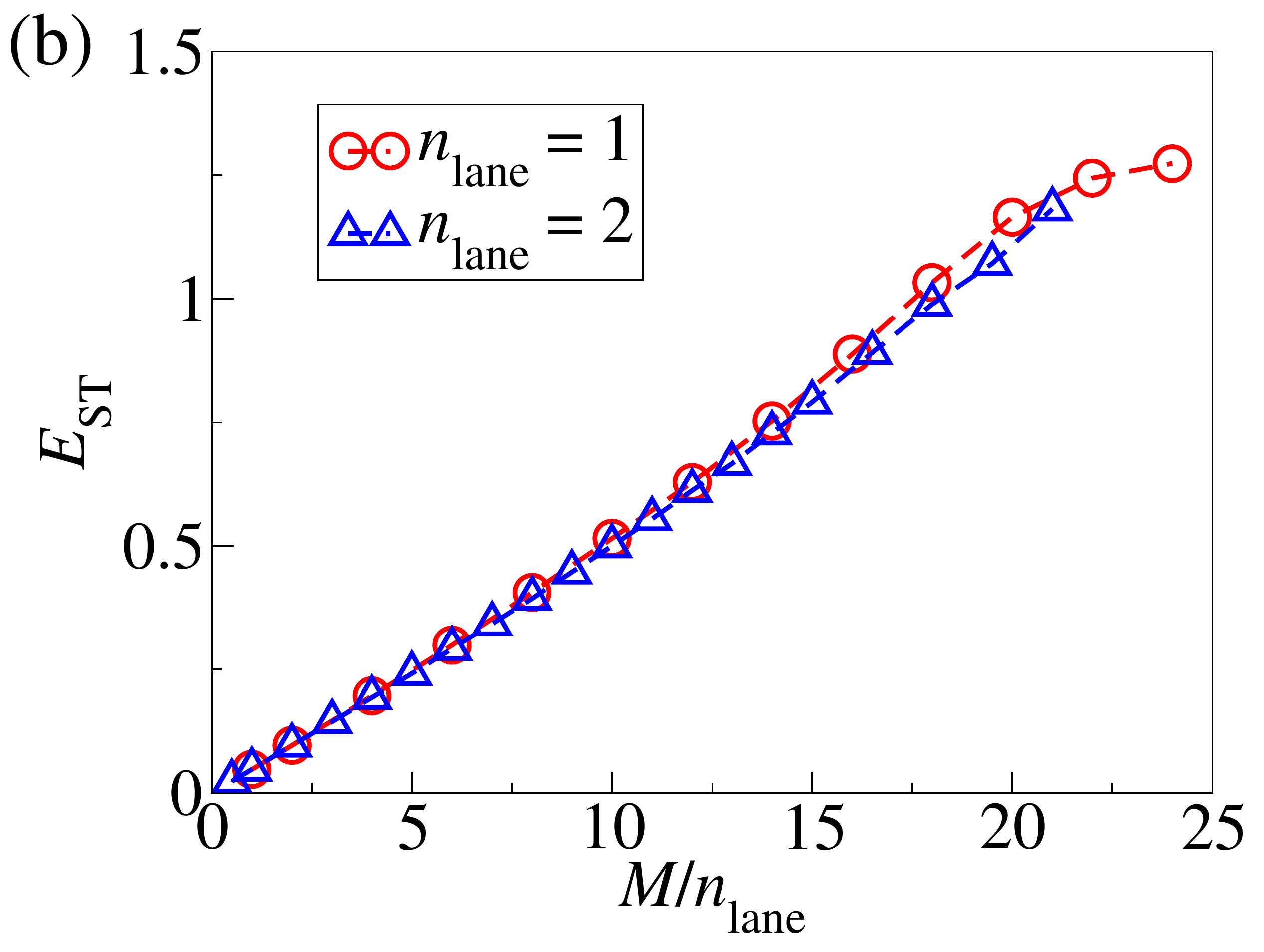, width=0.55\linewidth}
}
\caption{\red{(a) An example of a spatially replicated network from a random regular graph with $N=10$ nodes. Each red ellipse corresponds to a single node in the original spatial network, and consists of two \emph{sub-nodes}. All sub-nodes and links in the modified network are subject to constraint Eqs. (\ref{eq_con2}) - (\ref{eq_con4}) as for nodes and links in single-lane networks. Although there are four connections between two neighboring ellipses, at most two vehicles can pass through any two of the four directed links, since each sub-node can be occupied by at most one vehicle. Hence, the modified network corresponds to a network with two lanes in each direction of traffic flow.
(b) The average optimized energy $E_{\rm ST}$ as a function of $M/n_{\rm lane}$, i.e. the number of vehicles rescaled with the number of lanes $n_{\rm lane}$, for both cases of single-lane and double-lane networks. The original spatial networks are random regular graphs with $N=100$ and $K=3$.}}
\label{fig_multilane}
\end{figure}
%%%%%%%%%%%%%%%%%%%%%%%%%%

\red{In the main text, we considered networks where each node and directed link can be occupied by at most one vehicle, as constrained by Eqs.~(\ref{eq_con3}) and (\ref{eq_con4}). It corresponds to cases where every road is bi-directional, but is single-lane in each direction. To generalize our derived results to transportation networks with multiple lanes in each direction, one can re-derive all the equations in the conventional cavity approach and the ST algorithm to allow more than one vehicles on each directed link. Nevertheless, this approach may greatly increase the computational complexity of the derived algorithms.}

\red{
A simple alternative approach is to modify the topology of the original spatial network to accommodate multiple lanes. As an example, we consider networks where each node and directed link can be occupied by two vehicles. As shown in \fig{fig_multilane}~(a), we replicate every node and connect every original and replicated node to their original and replicated neighbors. Each ellipse in \fig{fig_multilane}~(a) represents a single node in the original network, and consists of two \emph{sub-nodes} (i.e. the original and the duplicated nodes). As we can see, there are four connections between two neighboring ellipses, which corresponds to eight directed links with four of them in each direction. Nevertheless, since each sub-node can be occupied by at most one vehicle, at most two vehicles can pass through any two of the four directed links between neighbor ellapses. We then convert this spatially replicated network into a space-time network, and employ the ST algorithm to identify the optimal spatio-temporal path configurations on networks with multiple lanes.}

\red{
Compared with the case of single-lane networks considered in the main text, the capacity of the double-lane network to accommodate traffic flows should have been doubled. We therefore employ the ST algorithm to identify the optimal path configurations on networks with two lanes in each direction. The results of average optimized energy $E_{\rm ST}$ as a function of $M/n_{\rm lane}$ for both cases of single-lane and double-lane networks are shown in \fig{fig_multilane}~(b), whereas $n_{\rm lane}$ denotes the number of lanes. The collapse of the results from the two cases suggest that the capacity of the double-lane network has roughly doubled. These results also suggest that the ST algorithm is able to identify the optimal spatial-temporal path configurations on networks with multiple lanes, by first replicating the spatial network and then converting it to a space-time network.
}

%%%%%%%%%%%%%%%%%%%%%%%%%%
%\begin{thebibliography}{0}
%\end{thebibliography}

\vspace{-0.5cm}
%\bibliographystyle{h-physrev4}
%\bibliography{network19}

\end{document}